\documentclass[12pt]{article}
\usepackage{natbib}
\usepackage{epsfig}   
\usepackage{epstopdf} 
\usepackage{graphicx}
\usepackage[a4paper]{geometry}
\newcommand{\dd}{\,\mathrm{d}}

\begin{document}

\begin{center}

A Single Physical Model for Diverse Meteoroid Data Sets\\

\bigskip\bigskip\bigskip

Valeri V.\ Dikarev,$^{a,b,\ast}$
Eberhard Gr\"un,$^{a,c}$
William J.\ Baggaley,$^d$\\
David P.\ Galligan,$^d$
Markus Landgraf,$^e$
R\"udiger Jehn$^e$\\

\bigskip\bigskip\bigskip

$^a$~Max-Planck-Institut f\"ur Kernphysik,\\Postfach 103980, 69029 Heidelberg, Germany\\
$^b$~Astronomical Institute, St.\ Petersburg University, 198504 St.\ Petersburg, Russia\\
$^\ast$~Corresponding author E-mail address: {\em Valeri.Dikarev@mpi-hd.mpg.de}\\
$^c$~Hawaii Institute of Geophysics and Planetology, University of Hawaii,\\1680 East West Road, Honolulu, HI 96822, USA\\
$^d$~Department of Physics and Astronomy, University of Canterbury,\\Private Bag 4800, Christchurch, New Zealand\\
$^e$~ESA/ESOC, Robert-Bosch-Strasse~5, 64293 Darmstadt, Germany\\

\bigskip\bigskip\bigskip
\end{center}

\noindent Pages:~\pageref{lastpage}\\
\noindent Figures:~\ref{lastfigure}\\

\newpage

\noindent {\bf Proposed Running Head:} A New Meteoroid Model\\

\bigskip\bigskip\bigskip

\noindent {\bf Editorial correspondence to:}\\
Dr. Valeri Dikarev\\
Max-Planck-Institut f\"ur Kernphysik\\
Postfach 103980\\
69029 Heidelberg\\
GERMANY\\

\noindent {\bf Please send packages to:}\\
Dr. Valeri Dikarev\\
Max-Planck-Institut f\"ur Kernphysik\\
Saupfercheckweg 1\\
69117 Heidelberg\\
GERMANY\\

\noindent Phone: 49-6221-516-543\\
Fax: 49-6221-516-324\\
E-mail address: {\em Valeri.Dikarev@mpi-hd.mpg.de}\\

\newpage

\noindent {\bf ABSTRACT}

The orbital distributions of dust particles in interplanetary
space are inferred from several meteoroid data sets
under the constraints imposed by the orbital evolution of the particles
due to the planetary gravity and Poynting-Robertson effect. Infrared observations
of the zodiacal cloud by the COBE DIRBE instrument, flux measurements
by the dust detectors on board Galileo and Ulysses spacecraft,
and the crater size distributions on lunar rock samples retrieved
by the Apollo missions are fused into a single
model. Within the model, the orbital distributions are expanded into a sum of
contributions due to a number of known sources, including the asteroid belt
with the emphasis on the prominent families Themis, Koronis, Eos and Veritas, as
well as comets on Jupiter-encountering orbits.
An attempt to incorporate the meteor orbit database acquired
by the AMOR radar is also discussed.

\bigskip\bigskip\bigskip

\noindent {\bf Key Words:} Meteoroids; Interplanetary Dust; Orbits

\newpage

\section{Introduction}

We construct a new meteoroid model to predict meteoroid
fluxes on spacecraft, with the revision of the earlier
models~\citep{Divine-1993,Staubach-1996,Staubach-et-al-1997}
being motivated by several reasons. First of all, a mistype in the computer program
has been known to affect the reduction of the Harvard Radio Meteor Project
(HRMP) data \citep{Taylor-1995}. This mistype alone
had tainted the previous meteoroid models,
since they relied on the reduced radar meteor distributions.
Moreover, \cite{Taylor-Elford-1998} pointed out
that the orbital distributions restored from the HRMP survey were
affected by yet another, unaccounted bias.
These authors argued that many high-speed meteors
remain unregistered because they ablate too high, too far from the radar,
and their echoes are too weak to detect.
Having corrected the mistype, Taylor and Elford reduced the HRMP data
again, accounting for the newly recognized bias as well.
However, their effort was undermined by missing observation
logs from the HRMP, so the reduction could not be perfected.
A new meteor survey was selected for incorporation in the
meteoroid model, the one conducted by the AMOR radar at Christchurch,
New Zealand, in the period from 1995 to 1999~\citep{Galligan-Baggaley-2004}.

Second, several other meteoroid data sets of high quality became
available for incorporation in the model. The COBE/DIRBE infrared
sky maps depicted the thermal emission of interplanetary dust (IPD)
as seen from Earth-bound observatory within a wide range of
solar elongations, through as many as 10 different filters from 1.25
to 240$\mu$m \citep{Kelsall-et-al-1998}.
The dust detectors on board Galileo and Ulysses
in deep space continued their operation
and collected new impact events worthwhile incorporation
in model as well.

Third, the expansion of computer memory allows one today
to detail the meteoroid distributions at a greatly improved
level using large multi-dimensional arrays. In particular,
the assumption of mathematical separability of the multi-dimensional
distribution in position and velocity of meteoroids
into single-argument functions of orbit elements
postulated in \citep{Divine-1993} and replicated since then,
is now partially lifted off.

Fourth, this new capacity of meteoroid model software
is exploited to replace the empirical separable distributions
of the previous models by the theoretical non-separable
distributions of meteoroids obtained via semi-analytical
dynamical simulations. These theoretical distributions,
constructed for a number of possible meteoroid sources,
including asteroids and comets, are then summed together
with the weights assigned to fit the observations.

In this paper, we concentrate on the constructed model description
and demonstration of its abilities to reproduce the incorporated
data sets as well as a few others, while a separate publication
is under preparation to cover various model's impications,
like the relative abundances and lifetimes of meteoroids from comets,
asteroids and interstellar dust in different regions of the Solar system.
In particular, Sect.~2 of this paper features
a new computer code that reads the multi-dimensional
distributions and produces the estimates of number densities,
fluxes and velocities of so distributed particles. In Sect.~3,
the dynamics of meteoroids of different sizes and origins are
revisited, semi-analytical models of their steady-state
distributions in orbit elements and mass are proposed.
Meteoroid data sets that are incorporated in the new model
are discussed in Sect.~4. The fitting of the simulated distributions
to the data is described and commented in Sect.~5.
The last Sect.~6 summarizes the results of modeling efforts.

\section{A new computer code for meteoroid models}

First applied in the engineering model by \cite{Divine-1993},
the concept of phase density of meteoroids is at the core of the new model, too.
The number of particles of reference mass per unit coordinate space and velocity space
volumes is given by function $n(x,y,z,\dot x,\dot y,\dot z)$,
assuming the standard notation for Cartesian coordinates.
The number of particles above a given mass~$M$ is obtained by
multiplying function~$n$ by the cumulative mass distribution~$H(M)$.

\cite{Divine-1993}, followed by \cite{Staubach-1996}, simplified the problem
by introducing three functions of a single argument,
$p_e(e)$, $p_i(i)$ and $p_q(q)$, that were related to the phase density via
\begin{equation}\label{Divine}
   n = {1\over2\pi e}\left({q\over GM_\odot}\right)^{3/2} p_q \times p_e \times p_i
\end{equation}
where $q$~is the perihelion distance, $e$~the eccentricity,
and $i$~the inclination of the particle orbit about the Sun.
The Solar gravitational parameter is denoted by~$GM_\odot$.
\cite{Matney-Kessler-1996} also provide the relationship between
$p_q$, $p_e$ and $p_i$ and the formally defined distributions in perihelion distance,
eccentricity and inclination $f(q,e,i)$
\begin{equation}
   f (q, e, i) = {2\pi^2\; q^2\; \sin i \over (1-e)^{3/2}} \; p_q \times p_e \times p_i
\end{equation}
which is straightforwardly related to the distributions in semimajor axis, eccentricity
and inclination
\begin{equation}
   f (a, e, i) = f(q,e,i)\; (1 - e).
\end{equation}
The equation~(\ref{Divine}) was postulated rather than derived, apparently
to simplify the integrands in the number density, flux and impact speed
expressions.

For the point of observation at distance~$r$ from the Sun and
latitude~$\lambda$ above the ecliptic plane, \cite{Divine-1993} introduced
auxiliary variables~$\chi=\arcsin q/r$ and $e_\chi=(r-q)/(r+q)$
and derived the spatial number density of the meteoroids above
the mass threshold~$M$
\begin{equation}\label{density}
   \rho = {H(M)\over\pi}
   \int_{\chi=0}^{\pi/2}
   \int_{e=e_\chi}^1
   \int_{i=|\lambda|}^{\pi-|\lambda|}
   {p_q \times p_e \times p_i \; \sin\chi \dd \chi \dd e \dd i
   \over (e-e_\chi)^{1/2} (\cos^2\lambda - \cos^2i)^{1/2}},
\end{equation}
as well as the impact frequency into dust detector (the limits of
integrals are as in the previous equation)
\begin{eqnarray}
   \nu &=& {1\over4\pi}\sum_{l=1}^4
   \int\!\!\int\!\!\int
   \Delta V_l \; A [\Delta \vec V_l] \; H (M_{\mathrm{min}} [\Delta V_l]) \times\nonumber\\
   & &
   \phantom{{1\over4\pi}\sum_{l=1}^4\int\!\!\int\!\!\int}\times
   {p_q \times p_e \times p_i \; \sin\chi \dd \chi \dd e \dd i
   \over (e-e_\chi)^{1/2} (\cos^2\lambda - \cos^2i)^{1/2}},\label{flux}
\end{eqnarray}
and the average impact speed
\begin{eqnarray}
   \overline V &=& {\nu^{-1}\over4\pi}\sum_{l=1}^4
   \int\!\!\int\!\!\int
   \Delta V_l^2 \; A [\Delta \vec V_l] \; H (M_{\mathrm{min}} [\Delta V_l]) \times\nonumber\\
   & &
   \phantom{{1\over4\pi}\sum_{l=1}^4\int\!\!\int\!\!\int}\times
   {p_q \times p_e \times p_i \; \sin\chi \dd \chi \dd e \dd i
   \over (e-e_\chi)^{1/2} (\cos^2\lambda - \cos^2i)^{1/2}}.\label{speed}
\end{eqnarray}
In the equations above, $\Delta \vec V_l = \vec V_l - \vec V_{\mathrm{det}}$,
$\vec V_l = \vec V_l (r, \lambda; q,e,i)$ is the heliocentric velocity of meteoroids
approaching the dust detector located at $(r,\lambda)$ from the $l$-th direction out of four
(inward and outward, ascending and descending motion of particles in orbits with given $q$, $e$, $i$),
$\vec V_{\mathrm{det}}$ is the heliocentric velocity of detector, $A(\Delta\vec V)$ is the detector's
sensitive area exposed to the particle flux along~$\Delta\vec V$. The minimum mass of meteoroid
that can be detected at the impact speed $\Delta V$, $M_{\mathrm{min}}[\Delta V]$,
characterizes the detector sensitivity.

In the new code for meteoroid models, the product of the three functions of a single argument
$p_q \times p_e \times p_i$ is first replaced by a single function of three arguments $D(q, e, i)$.
The function~$D$ is discretised and stored in a three-dimensional array allocating
50 bins over the perihelion distance range of 0.05 to 6~AU, 100 bins over the eccentricity
range from 0 to 1 and 180 bins over the inclination range from 0 to $\pi$.
The perihelion bins are spaced logarithmically.

In order to eliminate redundant computations under the circumstances when
several meteoroid populations must be processed simultaneously, the function~$D$
is turned into a vector function, $\vec D$, composed of the scalar functions
standing for sole populations. The integrals~(\ref{density}) and~(\ref{flux}) are then
implemented over vector functions adopting usual linear algebra operations
with vectors and scalars, but with an unusual definition of the
product of two vectors, which is done componentwise, yielding vector of the same dimension again.
The result is a vector of number densities~$\vec\rho$ or impact frequencies~$\vec\nu$
due to all meteoroid populations. A componentwise vector inversion is also defined
for the average impact speed calculations in~(\ref{speed}) where the inverse
frequency~$\vec\nu^{-1}=(\nu_1^{-1},\ldots,\nu_N^{-1})$ is required.

The code is designed to work with the Solar radiation pressure to gravity ratios~$\beta$
of meteoroids ranging from 0 to 1.
The expressions for the heliocentric velocity of meteoroids
can be found both in \citep[][for $\beta=0$]{Divine-1993} and
\citep[][for $0\leq\beta<1$]{Gruen-et-al-1997}.
The case $\beta=1$ is accounted for by the monodirectional
streams of particles flowing through the Solar system introduced below.
When the ratio is same for all populations,
like in the Divine model ($\beta=0$), then the heliocentric velocity
of meteoroids required in (\ref{flux}) and (\ref{speed}) can be calculated
once for all populations. When the ratios are different, like in the Staubach
model with two populations of interplanetary dust having $\beta=0$,
two more having $\beta=0.3$, and the last one having $\beta=0.7$, then
the velocities of meteoroids are different, too.
Nevertheless, even in this more complicated case,
the directions of meteoroid velocities relative to the Sun
can be calculated once for all $\beta$~ratios, and then
their magnitudes are scaled in accord with their $\beta$~ratio.

The approximate numerical evaluation of integrals is done using the quadrature
formulae of the highest algebraic precision due to Gauss and Chebyshev-Gauss
for each of the nested one-dimensional integrals. The quadrature formulae allow for
less evaluations of integrands in order to reach a given accuracy, the quadratures
are also crucial to eliminate the singular expressions in the integrands from
the actual computations: they are isolated in the weight functions
of the quadrature formulae, and the rest integrands that are evaluated numerically
do not contain anything that could lead to overflows.

Several optimisations have been introduced to further speed up the computations
in special cases. Although the code is capable to process the three-dimensional
functions~$D$, it can also be run with the previous models composed of the
products $p_q \times p_e \times p_i$. In this case, the
inclination integral can be separated from the two other integrals in the
number density integral~(\ref{density}), leading to a significant gain
in computation speed.

When the code is applied to predict fluxes or number densities
of the meteoroids above fixed mass thresholds, then the cumulative mass
distribution~$H(M)$ can be calculated once for all populations,
all velocity vectors, simplifying the integrand and thus accelerating
the computations. When more complicated selection criteria should be met,
involving combinations of the mass and impact speed of meteoroids,
then the full computation is done.

The interstellar dust in the Solar system is represented in a very
simplified form only, the one adopted in the previous meteoroid model
by \cite{Staubach-1996}. A monodirectional stream of interstellar grains is
parameterised by its downstream direction, flow speed, and the cumulative mass
distribution of the constituent particles. Since a considerable fraction
of the interpanetary dust observed by Galileo and Ulysses has the $\beta$ ratio
close to unity, a monodirectional stream is a good approximation, especially
at longer distances from the Sun.
Multiple streams can be processed simultaneously, however,
along with multiple populations of dust on bound orbits.

\section{Meteoroid sources, dynamics and distributions}

Unlike the previous models by Divine and Staubach, populations
of interplanetary meteoroids in the new model are not empirically
derived from the observations alone. A theoretical model
of generation and dynamical distribution of particulate matter
from a number of prominent sources is developed to complement scarce data.
A simple view upon the sources of dust and the forces distributing
it over the Solar system is adopted. Dust particles of all sizes are
assumed to be produced via collisional destruction of larger boulders
in the asteroid belt, and on the Jupiter-crossing orbits
where most of the comets reside.

In the new meteoroid model, the mass distribution is separated from
the orbital distributions, a realistic assumption over wide mass ranges.
The model's mass distribution of meteoroids is based on
the following considerations. In the collisional destruction experiments,
the number of the fragments greater than~$M$ in mass was found to obey the power law
\begin{equation}
   H^+(M) \propto M^{-\gamma}
\end{equation}
with indeces~$\gamma$ belonging to the range from 0.6 to 0.9~\citep{Gruen-et-al-1980,Asada-1985}.
This law describes the dust production rate as well in continuous collisions
between meteoroids. The time the meteoroids spend in a given orbital space bin $T^-$
is determined by the removal process. The equilibrium number
of particles in orbital space bin is then
\begin{equation}
   H(M) = H^+(M) \times T^-.
\end{equation}

According to \cite{Gruen-et-al-1985}, the particles bigger than $\sim10^{-5}$~g
have cross-section area sufficiently large to make the collisional destruction
by the smaller particles the dominant removal mechanism.
Due to the Poynting-Robertson effect, the particles smaller than $\sim10^{-5}$~g
are typically evacuated from the orbital space bin where they were produced,
before they can collide.

The rate of removal depends on meteoroid mass in both scenarios.
\cite{Ishimoto-2000} proposed an illustrative explanation
of the mass distribution of meteoroids observed at 1~AU
(in Sect.~3.2 of his paper).
The strength of the Poynting-Robertson effect and the induced drift rates
of particles away from their origin are proportional to the inverse particle size,
thus the dwell time near the origin is proportional to the particle size,
i.e., $T_{\rm P-R}^-\propto M^{1/3}$.
The lifetime against collision with a significantly smaller projectile,
assuming no dependence on the projectile size, is inversely proportional
to the target particle area, i.e., $T_{\rm C}^-\propto M^{-2/3}$.
Combining these lifetimes with the collisional fragment mass distribution
with a plausible index $\gamma=2/3$, \cite{Ishimoto-2000} obtains
a distribution function very similar with the flux at 1~AU by \cite{Gruen-et-al-1985}.

There is a little problem with this illustrative explanation, however,
since in the regime of collisional destruction it does not account
for the fact that the mass of the disrupting projectile is roughly proportional
to the mass of the target, so that the smaller targets
can be destroyed by smaller projectiles that are more abundant,
collisions with them are more likely than in the framework of
simple mono-size projectile illustration provided in
\citep{Ishimoto-2000}.

\begin{figure}[h]
\centerline{\psfig{figure=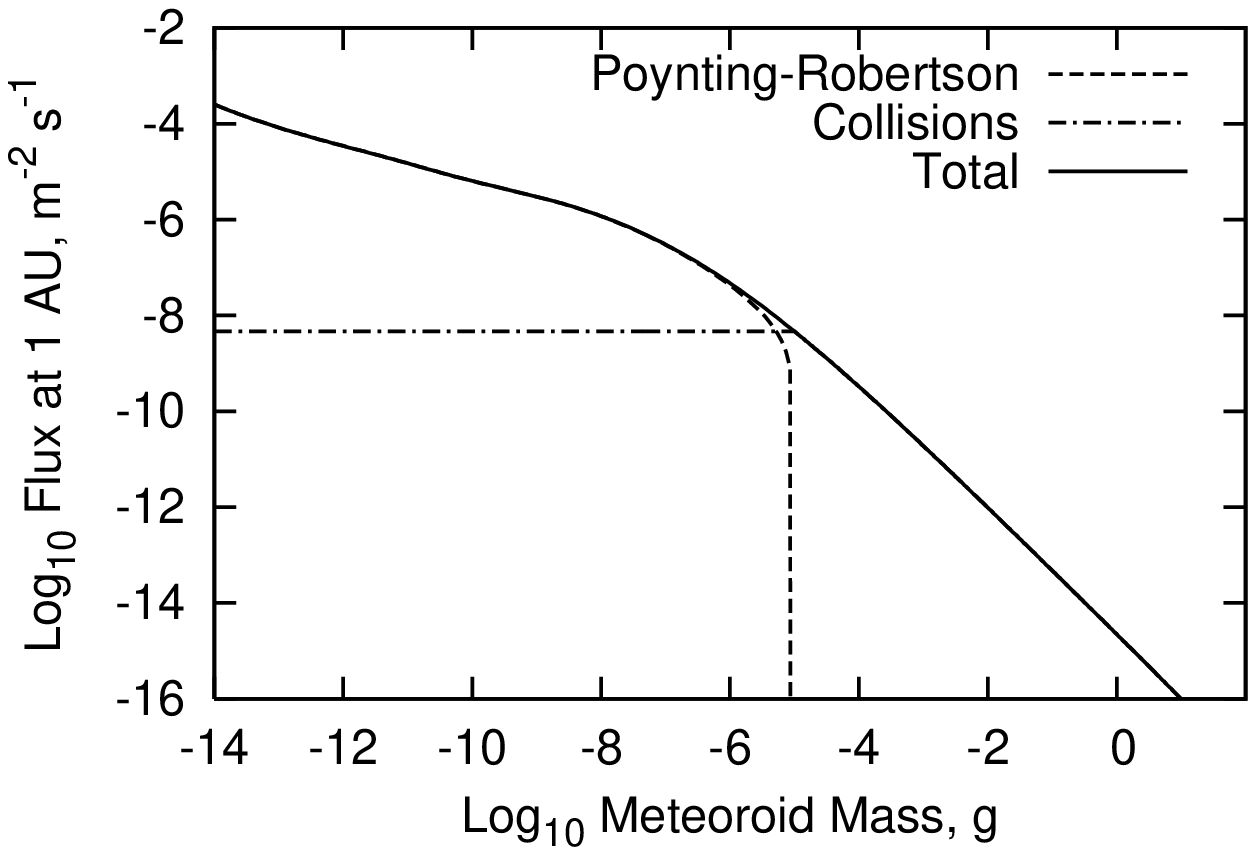,width=0.6\hsize,angle=270}}
\caption{The cumulative mass distributions of interplanetary meteoroids adopted in the new model.\label{Mass Distrib}}
\end{figure}

Therefore, the slope of mass distribution of meteoroids in the collisional
regime requires a more elaborate theory to be explained.
In the new meteoroid model the mass distribution $H(M)$ is postulated
rather than derived, based on the cumulative mass distribution of meteoroid
flux at 1~AU \citep{Gruen-et-al-1985}. It is reproduced in Fig.~\ref{Mass Distrib}.
The distinction between the dynamical regimes, the Poynting-Robertson drift
and collisional destruction at the origin,
is still implemented in the model since the orbital distributions of meteoroids
in the two regimes are drastically different. In what follows, the mass distribution
in the Poynting-Robertson regime (dashed curve in Fig.~\ref{Mass Distrib}) is
designated by $H_{\rm P-R}(M)$, and for that in the collisional regime the notation~$H_{\rm C}(M)$
is used (dash-dotted curve in Fig.~\ref{Mass Distrib}).

%
%
In the asteroid belt, the dust production rate is defined to be
proportional to the quantity of numbered asteroids.
We took number~(1) through~(13902) as provided by the Minor Planet Center
(the MPCORB database). For the sake of simplicity, equal production efficiencies of all
parent bodies were assumed. The great quantity of the asteroids as well as
their confinement to low inclinations and eccentricities
allow one to generate the distributions of good statistical quality
by simply counting the objects in orbital space bins.

\begin{figure}[h]
\psfig{figure=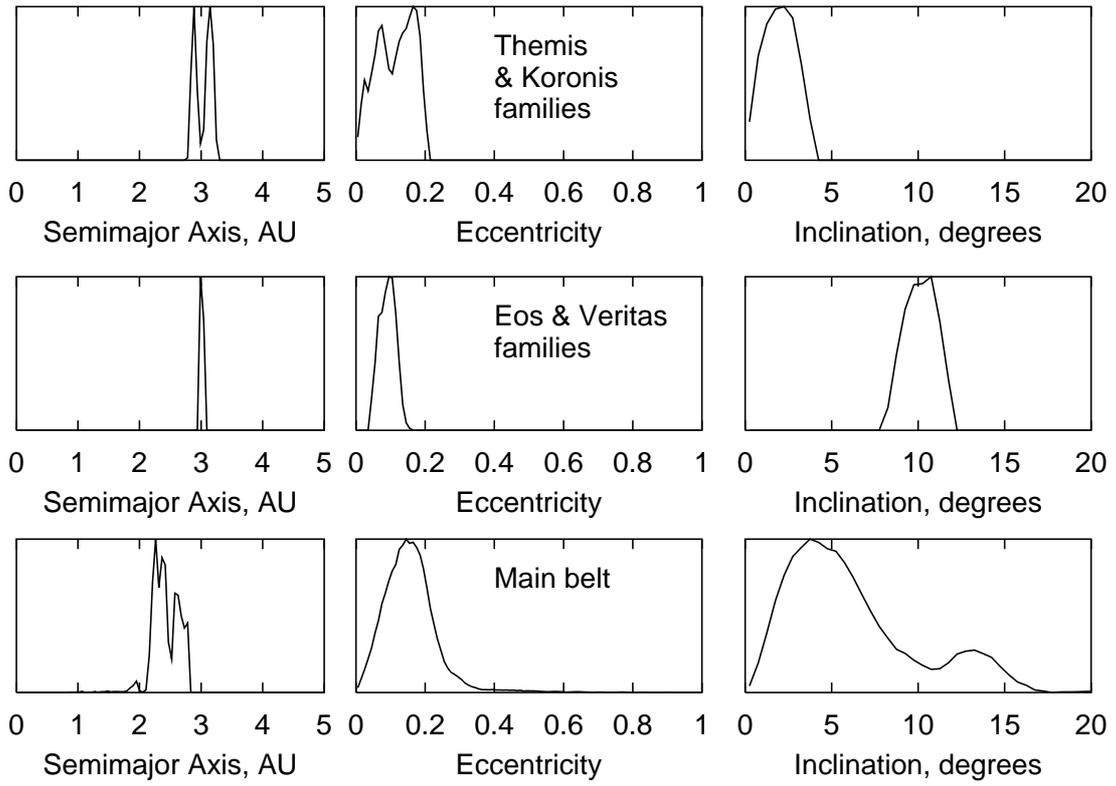,width=0.7\hsize,angle=270}
\caption{The orbital distributions of asteroids and big meteoroids ($M>10^{-5}$~g).\label{Asteroidal Collisional}}
\end{figure}

The result of this operation is shown in Fig.~\ref{Asteroidal Collisional}.
The three-dimensional distributions are integrated over
two arguments in order to produce comprehensive
plots. \cite{Dermott-et-al-1984} discovered the asteroid dust bands
extending from several asteroid families toward the Sun.
In order to allow the families to play a role in the new meteoroid
model, three distinct populations are recognized in the asteroid belt,
the Themis and Koronis families ($2.8<a<3.25$~AU, $0<e<0.2$, $0<i<3.5^\circ$),
Eos and Veritas families ($2.95<a<3.05$~AU, $0.05<e<0.15$, $8.5^\circ<i<11.5^\circ$),
and the main belt ($a<2.8$~AU). The Themis and Koronis
families are rather close in orbital space and are not separated
in the meteoroid model, so are the Eos and Veritas families,
both being associated with the ten-degree band
\citep{Grogan-et-al-1997,Dermott-et-al-2002}.
We denote the orbital distributions of the main-belt asteroids by $g_{\rm MB}(a,e,i)$,
of the Themis and Koronis families by $g_{\rm TK}(a,e,i)$, and of the Eos and Veritas families
by $g_{\rm EV}(a,e,i)$.

The orbital distributions of the three population were combined with the mass distribution
in the collisional regime $H_{\rm C} (M)$, reflecting
the fact that the big meteoroids are destroyed at their origin
before their orbital distributions evolve,
yet at the rates depending on the meteoroid sizes.

%
%

The small particles from asteroid collisions and, even more importantly,
from disintegration of the big meteoroids in the collisional regime,
are assumed to be distributed by the Poynting-Robertson effect.
\cite{Gorkavyi-et-al-1997} provide the solution for the orbital density
of dust evolving under the pure Poynting-Robertson effect
\citep[found earlier by][]{Leinert-et-al-1983}
\begin{equation}\label{WW int}
   f(a,e,i)\; e^{1/5} \sqrt{1-e^2} = \mathrm{const}_1
\end{equation}
as well as the one-dimensional distribution in semimajor axis for a point source
that is more practical to use
\begin{equation}\label{Leinert int}
   f(a) {2+3e^2\over a(1-e^2)^{3/2}}=\mathrm{const}'_1
\end{equation}
along any trajectory defined by the integral \citep{Wyatt-Whipple-1950}
\begin{equation}
   {a (1 - e^2) \over e^{4/5}} = \mathrm{const}_2,
\end{equation}
with the inclination~$i$ staying constant.

\begin{figure}[h]
\psfig{figure=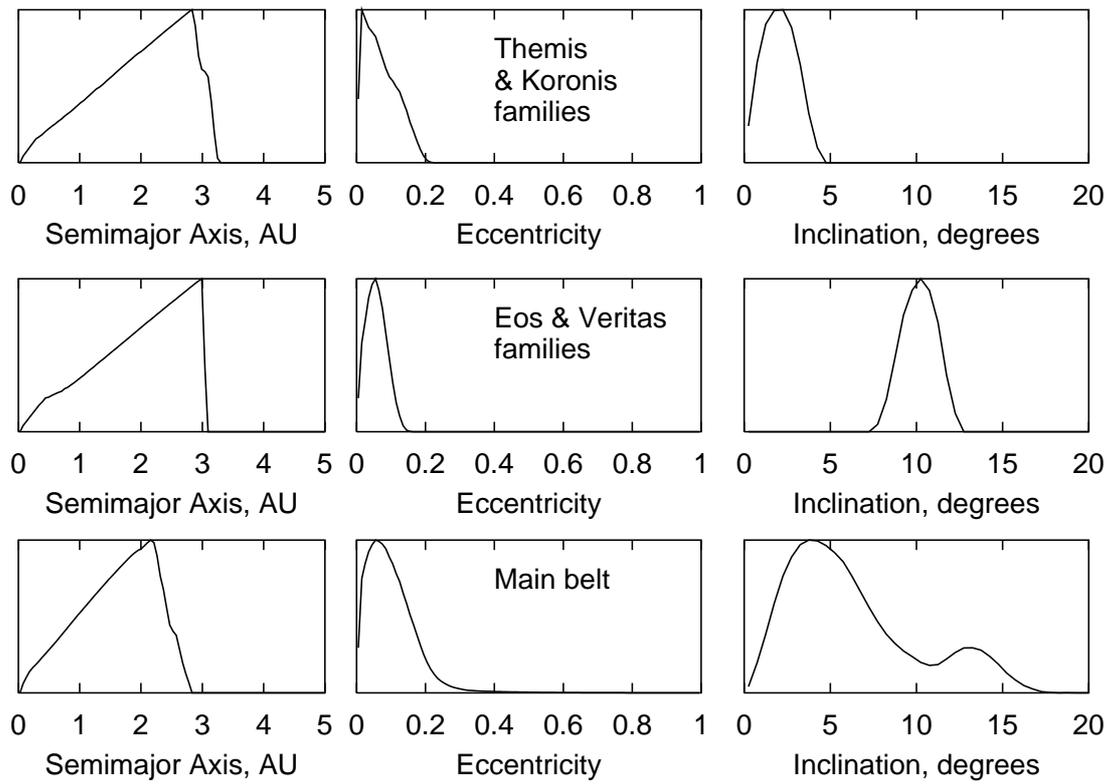,width=0.7\hsize,angle=270}
\caption{The orbital distributions of small meteoroids ($M<10^{-5}$~g) from asteroids spiraling
toward the Sun under the Poynting-Robertson effect.\label{Asteroidal Poynting-Robertson}}
\end{figure}

This solution was used to calculate everywhere the orbital density that
at the origin was fixed to the orbital density of numbered asteroids, i.e.
to the same production rate as for the big meteoroids. However, the mass distribution
adopted in this case was naturally that of the Poynting-Robertson regime, i.e.~$H_{\rm P-R}(M)$.
Again, the mass distribution is separated from the orbital density
because the Poynting-Robertson effect leads all dust particles along the same trajectories,
yet at different drift rates. The orbital distributions of the small dust grains
from asteroids are plotted in Fig.~\ref{Asteroidal Poynting-Robertson}.
We denote the orbital distributions of dust from the main-belt asteroids by $f_{\rm MB}(a,e,i)$,
from the Themis and Koronis families by $f_{\rm TK}(a,e,i)$, and from the Eos and Veritas families
by $f_{\rm EV}(a,e,i)$.

%
%

The asteroids
are located in a calm region of the Solar system where the perturbations of their
orbits due to planetary gravity are relatively weak. The effect of planetary gravity
on a low-eccentricity, low-inclination meteoroid orbit in the asteroid belt
is simply precession, i.e., the longitudes of perihelion and node advance
at constant rates. The rates strongly depend on the semimajor axis of particle orbit.
Whenever there is a scatter of semimajor axes in a particle ensemble,
the planetary perturbations efficiently randomize the longitudes of orbits.
Already in the framework of two-body problem the Solar gravity alone causes
differential rotation of disks of particles---the mean anomalies are randomized,
too.

This means, actually, that the distribution of meteoroids in longitude of node,
argument of pericenter and mean anomaly are close to uniform. Note that this was one
of the assumptions made by Divine and Staubach---they neglected any dependence
of their distributions on the longitudinal angles. This assumption is left intact
in the new meteoroid model not only for the particles from the asteroid belt,
but also, for the sake of simplicity, for the particles from comets.

The orbital distributions of meteoroids from the comets on
Jupiter-crossing orbits can not be defined as easy as
the distributions of dust from asteroids.
Because of a number of loss mechanisms, such as ejection
from the Solar system by the giant planets and fading out, very few
comets are displayed at a time and listed in the catalogues.
Moreover, the catalogues are prone to observational biases
since the comet nuclei are revealed by gas and dust
shed at high intensity at the low perihelion
distances. The imperfect removal of these biases
and low-number statistics would have degraded the quality
of dust source distribution based on the catalogues.
The close encounters with Jupiter leading rapidly to
chaos in the orbital dynamics of comets and meteoroids,
erase meteoroid's ``memory'' of the parent body orbit,
mixing up the particles from different comets.
A dedicated model of the dust production and dynamics
on Jupiter-crossing orbits is therefore necessary,
independent on the catalogues of comets.

The numerical integration of the equations of the motion
of meteoroids has been done for a limited number
of parent comets. By simulating the orbital dynamics
of particles from comet 2P/Encke \cite{Liou-et-al-1995} demonstrated that
the comets are necessary to account for the full thickness
of the zodiacal cloud, with the dust from asteroids being
confined too close to the ecliptic plane. \cite{Liou-et-al-1999} computed
the trajectories of meteoroids of several sizes from comet
1P/Halley and its imaginary prograde clone. \cite{Cremonese-et-al-1997}
studied numerically the contributions of dust from comets
29P/Schwassmann-Wachmann and 26P/Griegg-Skjellerup to the inner zodiacal cloud.
\cite{Landgraf-et-al-2002} additionally simulated the orbital
evolution of meteoroids from the Edgeworth-Kuiper belt.
These objects were proven to be important sources of meteoroids yet
altogether they are far from the full range of observed dust producers.
\cite{Hughes-McBride-1990} presented results of
a very ambitious simulation of trajectories of meteoroids
(mass $>10^{-3}$~g) from 135 short-period
comets, placing 5000 particles in orbit of every parent comet.
However, they did not consider the long-period comets.
The meteoroid mass range scarcely covered by all numerical simulations
is another problem.

Ironically, the orbital distributions obtained by means
of numerical integration of test particle trajectories have
the drawback of comet catalogues, i.e.\ the low number
of objects to distribute. An attempt to build the four-dimensional
distribution in orbit elements and mass would result in either low
resolution or high noise. There is a method, however,
to derive the orbital distributions approximately
from several standard assumptions of statistical mechanics.

Assume the ergodic hypothesis holds true in the region of intensive
encounters with planet. Ignoring the planet's eccentricity,
the motion of test particle can be described in the framework
of the restricted circular three-body problem. The problem
admits one integral of motion, the Jacobi constant~$C_{\rm J}$
which at large distances from planet is closely represented
by the Tisserand quantity
\begin{equation}\label{T}
   T = a_{\rm J}/a + 2\sqrt{a/a_{\rm J}(1-e^2)}\cos i
\end{equation}
with $a_{\rm J}$ being the radius of the planet's orbit,
that is, in turn, can be replaced by the more intuitively clear
notion of the encounter speed with planet~$U$ measured in units
of planet's velocity relative to the Sun,
\begin{equation}
   U=\sqrt{3-T}.
\end{equation}
Simultaneously, the equations of the motion obey the Liouville theorem,
so that the phase volume is preserved along any trajectory.
The phase space is composed, however, of layers of $C_{\rm J}=\mathrm{const}$
to which any trajectory must be confined. Assumption of the ergodic hypothesis
then yields the phase density in suitable canonical variables
in the form $n=n(C_{\rm J})\approx n(U)$. Adopting the coordinates and
velocities of test particles in the Cartesian inertial system
centered at the Sun as canonical variables,
the canonical system of units in which the sum of the Solar and planetary gravitational
parameters is unit, and transforming the phase density into
the distribution in six Keplerian orbital elements by
applying the Jacobian
\begin{equation}
   {\sqrt{a} e \sin i \over 2} \nonumber
\end{equation}
one can write the distribution in the six elements of encountering
particles\footnote{To change to an arbitrary system of units,
multiply the Jacobian by $(GM_\odot)^{3/2}$.}
\begin{equation}\label{gaei6}
   g = n (U) I_{\rm E} \; {\sqrt{a} e \sin i \over 2}.
\end{equation}
Now all complexity of the problem of obtaining the orbital
distributions is contained in the determination of
function~$I_{\rm E}$ (of all six orbital elements, generally speaking)
which is equal to one on the subspace of phase space
where ergodicity is supposed, and zero otherwise.

The function~$I_{\rm E}$ is approximated to be unit if the heliocentric
orbit of particle crosses a certain torus along the planet's orbit,
i.e. when frequent close encounters with planet allow for hypothesizing
ergodicity, and zero elsewhere. The torus' radius is found to $\approx0.5$~AU
for Jupiter by comparing the function~(\ref{gaei6})
with statistical distributions obtained numerically.

The function~$n(U)$ is expanded into a family of step-functions
to distinguish between the populations of meteoroids
from the parent bodies contained in different spherical concentric
shells of encounter velocity~$\vec U$
\begin{equation}
   n(U) = \sum_{j=1}^{N_{\rm C}} n_j (U),
\end{equation}
where $n_j(U)>0$ in the range of encounter speeds from $[(j-1)/N_{\rm C}]\,U_{\rm max}$
to $[j/N_{\rm C}]\,U_{\rm max}$, where it specifies the source strength,
and zero elsewhere, with $U_{\rm max}=1+\sqrt{2}$ being the maximum
encouner speed still possible for a bound heliocentric orbit.

\begin{figure}[h]
\centerline{\psfig{figure=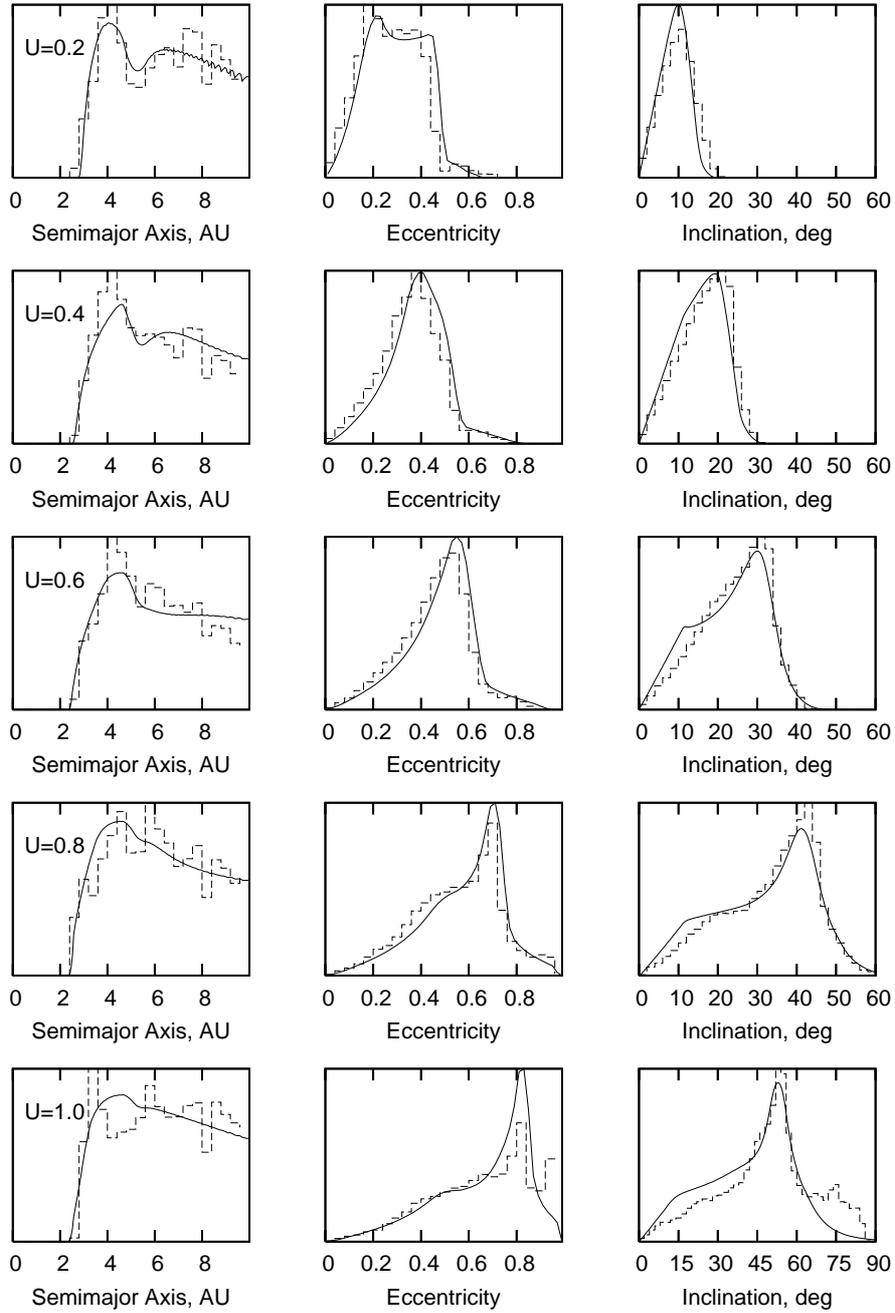,width=0.8\hsize}}
\caption{The orbital distributions of the particles scattered gravitationally by
a Jupiter-sized planet on a circular orbit 5.2~AU from the Sun.\label{gaei1}
Dashed-lined step-functions are the results of numerical experiments,
solid lines are the analytical solution.}
\end{figure}

\begin{figure}[h]
\centerline{\psfig{figure=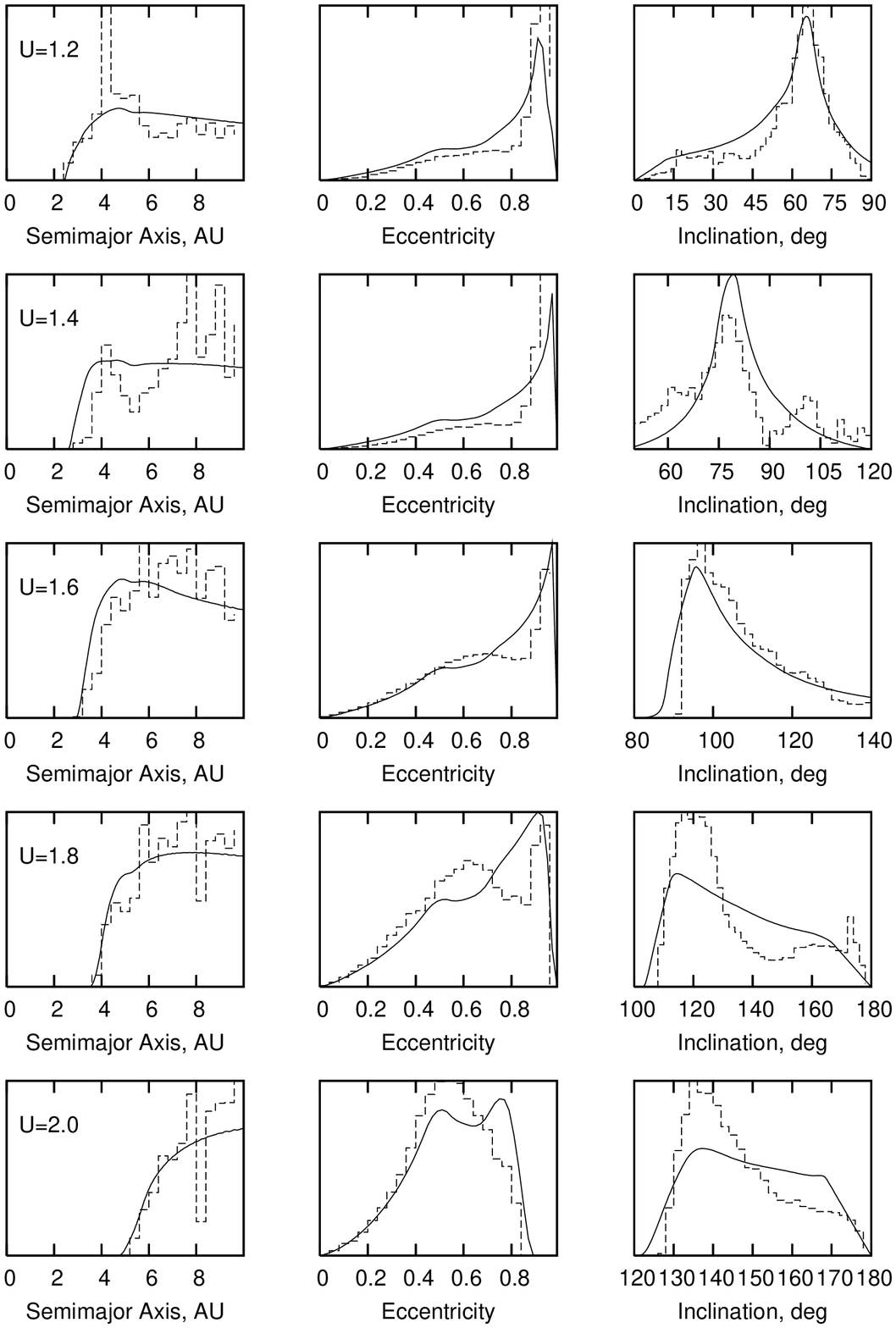,width=0.8\hsize}}
\caption{Same as Fig.~\ref{gaei1}\label{gaei2}}
\end{figure}

Thereby $N_{\rm C}=24$ distinct populations $g_j$ are recognized of
the parent objects in Jupiter-crossing orbits with arbitrary dust production
rates. Their orbital distributions were averaged over the longitudes
of node and pericenter to produce functions of only three arguments~$g_j(a,e,i)$
for the sake of compatibility with the number density
and flux calculation software developed for the new model.
Ten of them are shown in Fig.~\ref{gaei1} and~\ref{gaei2}
along with their numerically obtained counterparts.
Note that while the one-dimensional distributions in orbital elements
of simulated particles show low level of noise, facilitating the comparison
with analytics, the three-dimensional distribution would be unacceptably
distorted by the low-number statistics.

%
%

The numerical simulations were organized as follows.
In each experiment, one thousand massless particles
was injected into the system of Sun and Jupiter-on-a-circular-orbit
for every discrete value of encounter speed~$U$ from 0.2 to 2.0 step 0.2.
The encounter velocities were chosen randomly distributed
over the sphere of $U=\rm const$, the orbital angles
(the argument and longitude of perihelion, and the mean anomaly)
were also uniformly distributed random numbers initially.

The equations of the motion were solved by the MERCURY software package \citep{Chambers-1999},
using the energy-conserving Bulirsch-St\"or integration method.
The time of integration was 200,000 Jovian years, the output step 20 Jovian years.
All the intermediate positions were stored to accumulate statistics
of the orbital elements of the simulated particles. The results of numerical
simulations are ``mature'' since most of the particles that had not
been sorted out as resonant finished their orbital evolution---they were ejected
from the Solar system.

The orbits with semimajor axis spending too much time (set to 20,000~Jovian years, or one
tenth of the integration period) with non-jumping (staying within any 0.5~AU-wide range)
semimajor axis below 10~AU were sorted out, however. The constancy of semimajor axis
during that long period is very improbable for a particle with the orbital angles uncorrelated
with Jovian anomaly. The maximum \"Opik time of collision with a scattering sphere around
Jupiter of the radius~$0.5$~AU (our torus section radius) is below 10,000~Jovian years
even for the longest semimajor axis in the range of interest, 10~AU.
An encounter-preventing correlation of the orbital angles, on the other hand,
implies a resonance.

Each resonance is a potential breaker of the ergodicity assumption,
since even bound periodic orbits are possible in resonance. A dedicated theory
of the particle motion near resonances should be applied to describe
the resonant clouds in the restricted circular three-body problem
and their contribution to the overall distribution function.
The resonant populations are not included in the present model.

The radiation pressure and the Poynting-Robertson effect
are negligible perturbations with respect to the effect
of close encounters with Jupiter down to the one-micrometer-sized
meteoroids. Therefore, the orbital distributions~$g_j(a,e,i)$ can be
used to describe the populations of dust grains of all masses
above $\sim10^{-12}$~g. Nevertheless, the mass distributions
$H_{\rm P-R}(M)$ and $H_{\rm C}(M)$ are combined with $g_j$ below and above $10^{-5}$~g,
respectively.

The rationale for the division into $H_{\rm P-R}$ and $H_{\rm C}$
in the gravity-dominated region is as follows. The orbital evolution in
the regime of close encounters with Jupiter is very rapid,
so it should be capable to compensate for the loss of dust
at the inner boundary of the region of encounters
into the encounter-free zone instantaneously,
yet at the expense of the deep layers of the region.
Thus the dust in the region of encounters should
also be lost at the rate influenced by the Poynting-Robertson drag.
Although more loss mechanisms act concurrently in the region of encounters,
notably the ejections from the Solar system,
there is no observational evidence the mass distribution
is strongly different there from that of the flux at 1~AU.

As some particles pass into the internal encounter-free region of the Solar system,
they move under the Poynting-Robertson effect
in accord with Eq.~(\ref{WW int}), with their orbital density being
described by Eq.~(\ref{Leinert int}). Since their origin is in the region
of Jupiter-crossing orbits, it is reasonable to bind their
distributions to certain production rates there, and, analogous to the
parent bodies in Jupiter-crossing orbits, to introduce distinct
populations of the leaking particles categorized by the encounter speed
at Jupiter at the time of the crossing of the inner boundary of the region
of encounters set at $4.7$~AU.

\begin{figure}[h]
\centerline{\psfig{figure=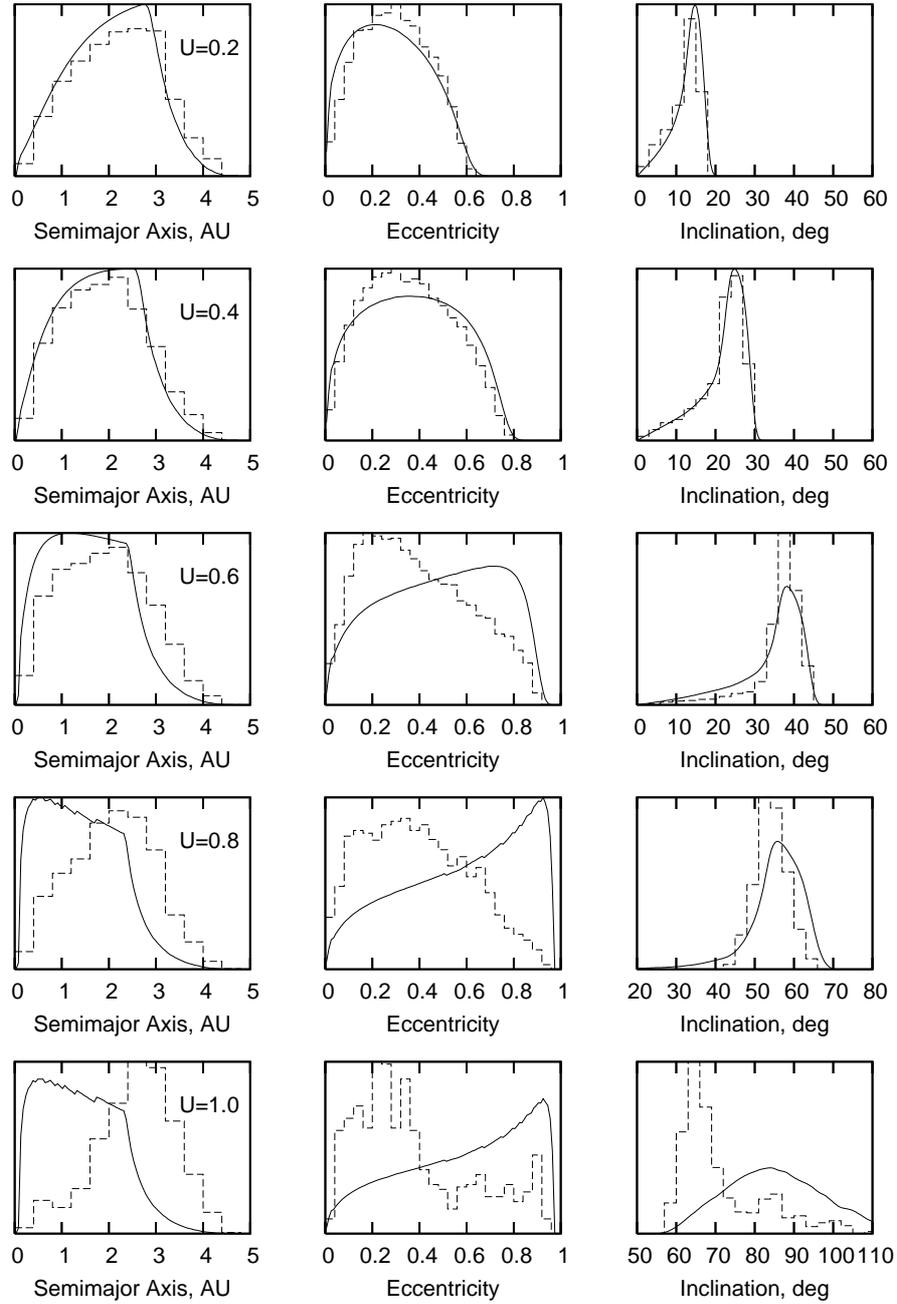,width=0.8\hsize}}
\caption{The orbital distributions of the particles of a radius 10~$\mu$m
leaking from the region of close encounters with Jupiter
after being multiply scattered by this planet's gravity.
The gravity of major planets excepting Mercury and Pluto is taken into account.
The solid curves show the analytics, the dashed-line step-functions are numerically
accumulated statistics.\label{faei1}}
\end{figure}

\begin{figure}[h]
\centerline{\psfig{figure=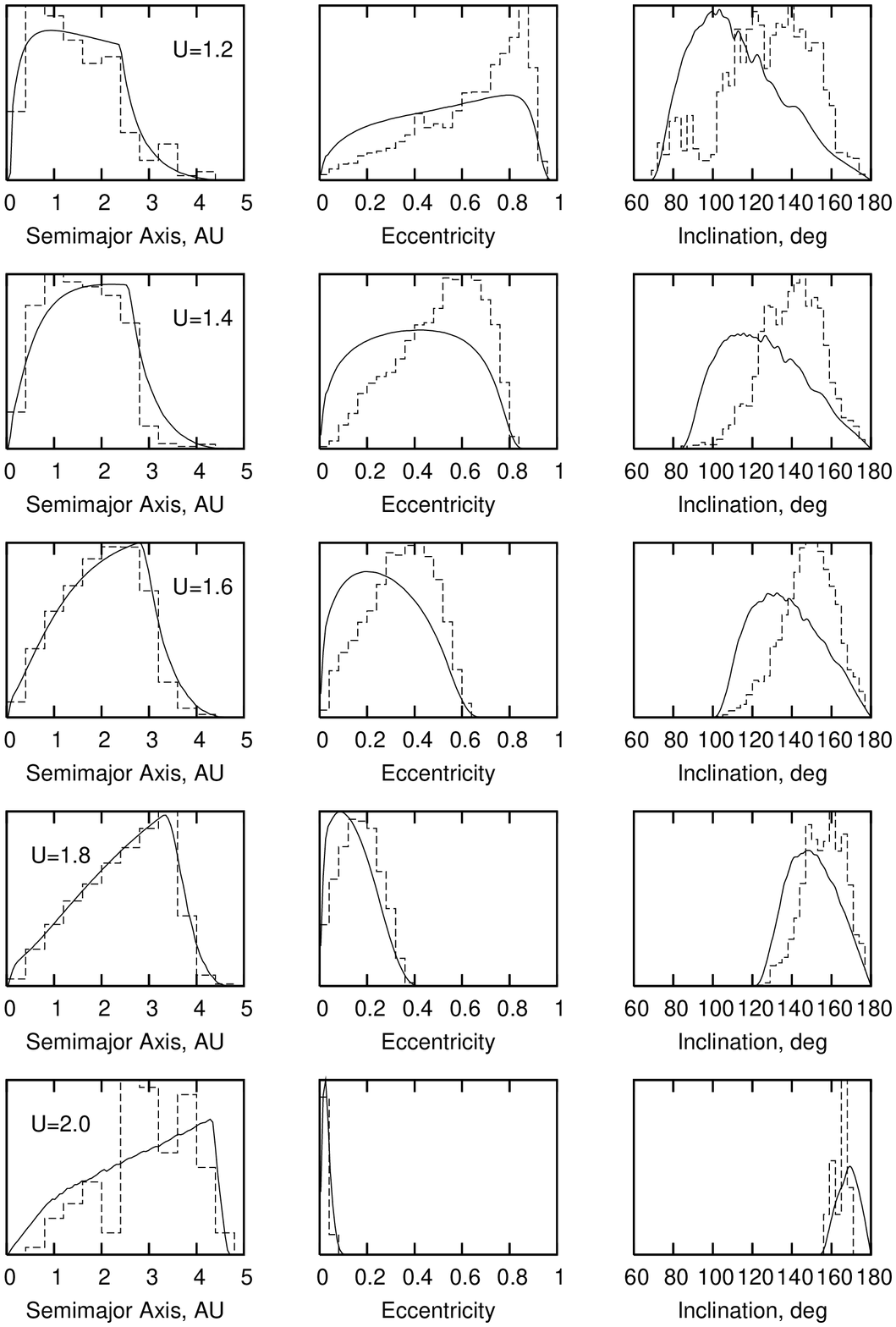,width=0.8\hsize}}
\caption{Same as Fig.~\ref{faei1}.\label{faei2}}
\end{figure}

The orbital distributions $f_j(a,e,i)$ of the leaking dust grains
escaping the region of encounters with $j=1,\ldots,N_{\rm C}$
adjacent values of $U$ are shown in Fig.~\ref{faei1} and~\ref{faei2}
along with their numerically obtained counterparts. In the model,
they are combined with the Poynting-Robertson mass distribution~$H_{\rm P-R}(M)$.

The numerical simulations to test $f_j(a,e,i)$ were set up different from the pure
gravitational problem. In each experiment, thousand particles
having radii of 10~micrometers were injected into the system
of Sun and seven major planets (Venus through Neptune) having
their real orbit elements, for every value
of the encounter speed from 0.2 to 2.0 step 0.2.
The orbit elements of test particles were generated
randomly as in the previous case.

The equations of the motion were solved by a Gauss-Radau code \citep{Everhart-1985}.
The period of integration was $5\cdot10^5$ years, the output step $10^3$ years. The intermediate positions were
stored to accumulate statistics of the orbital elements of the simulated particles.
Most of the particles finished their orbital evolution---they
were ejected from the Solar system or absorbed by the Sun.
No measure has been taken to mitigate resonances in this case.
The approximate analytical approach gives realistic distribution
functions for all encounter speeds except near and moderately below
$U=1$~Jovian velocity about the Sun.
Numerical simulations were also performed for the particles
of radii 3, 30, and 100 micrometers, showing a similar agreement
with the analytics, although statistics
were worse in the case of the bigger meteoroids since their
orbital evolution is slower and fewer trajectories could be obtained.

%
%

\begin{figure}[h]
\centerline{\psfig{figure=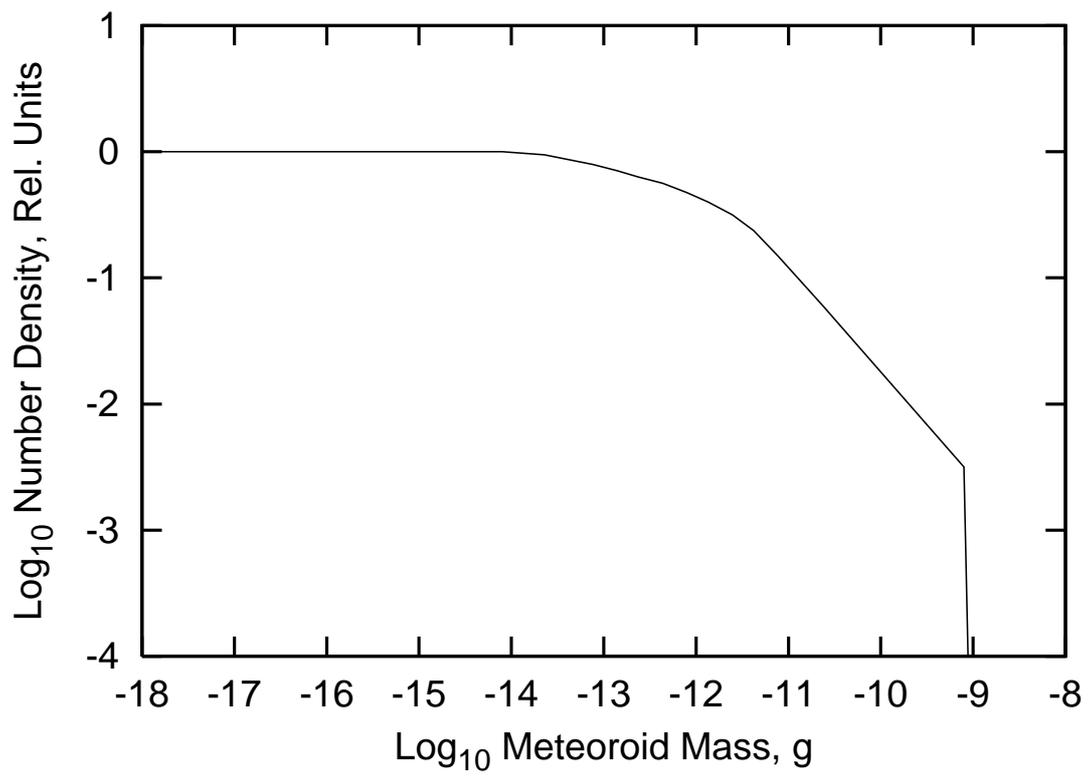,width=0.7\hsize,angle=270}}
\caption{The cumulative mass distribution of interstellar dust adopted in the new model.\label{ISD mass}}
\end{figure}

The interstellar dust in the Solar system is represented by a monodirectional
stream of particles, employing the approach of \citep{Staubach-1996}.
However, the cumulative mass distribution was initially changed to the
more recently derived curve of \citep{Landgraf-et-al-2000},
and then, responding to the new calibration of the DDS instrument
(see Sect.~\ref{G&U DDS}), the latter was shifted along the mass scale.
The relative mass distribution adopted for the meteoroid
model can be inspected in Fig.~\ref{ISD mass}.
The downstream direction of the interstellar dust stream is set
to 77$^\circ$ ecliptic longitude and $-3^\circ$ ecliptic latitude,
and the speed relative to Sun is set to 26~km~s$^{-1}$,
to match the parameters of the neutral gas flow
through the Solar system \citep{Witte-et-al-1993}.
The absolute normalization of the ISD flux was left
free in order to allow for a new balance between the
interplanetary and interstellar dust in the Galileo
and Ulysses data, after a re-formulation of the interplanetary
meteoroid populations.

In total, six populations of dust from asteroids are introduced in the model,
72~populations of dust from comets and other parent objects on
Jupiter-crossing orbits, and one stream of interstellar dust (ISD),
with the total orbital density of objects above a given mass specified by
\begin{eqnarray}
   x_1 \, (H_{\rm C} \times g_{\rm MB} + H_{\rm P-R} \times f_{\rm MB}) \;+\nonumber\\
   x_2 \, (H_{\rm C} \times g_{\rm TK} + H_{\rm P-R} \times f_{\rm TK}) \;+\nonumber\\
   x_3 \, (H_{\rm C} \times g_{\rm EV} + H_{\rm P-R} \times f_{\rm EV}) \;+\nonumber\\
   \sum_{j=1}^{N_{\rm C}} x_{3+j} H_{\rm C} \times g_j                \;+\nonumber\\
   \sum_{j=1}^{N_{\rm C}} x_{3+N_{\rm C}+j} H_{\rm P-R} \times g_j      \;+\nonumber\\
   \sum_{j=1}^{N_{\rm C}} x_{3+2N_{\rm C}+j} H_{\rm P-R} \times f_j     \;+\nonumber\\
   x_{3+3N_{\rm C}+1} \times \mbox {ISD}.\phantom{\;+}
\end{eqnarray}
However lengthy it is, this expression poses a linear inverse
problem for the weights $x_k$ which is still much easier
to solve than the non-linear problems of \citep{Divine-1993,Staubach-1996}.

The populations of meteoroids ($H_{\rm C}(M)$)
and dust grains ($H_{\rm P-R}(M)$) from asteroids are required to have
the same normalization factors though, as the mass distribution at the source is
expected to be naturally smooth. In contrast,
the populations of meteoroids and dust from comets are not a-priori
synchronized in their normalization even when they
correspond to the same encounter speed~$U$.
No assumption is put here to allow for more flexibility of fit.
The effects that shape the mass distributions of particles
on Jupiter-crossing orbits are rather difficult to simulate numerically,
and we have not proven any particular synchronization to be applied
in the model.

%
%

\section{Meteoroid data sets}

Already \cite{Divine-1993} incorporated in his meteoroid model a number of
data sets obtained by different observation methods.
It was the interplanetary flux model by \cite{Gruen-et-al-1985}
embracing several early data sets, with the most significant
being the microcrater counts on lunar rock samples retrieved by
the Apollo missions, meteoroid impact records by the dust detectors on board Pioneer~10 and 11,
Helios, Galileo spin-averaged fluxes measured between 0.88 and 1.45~AU,
Ulysses spin-averaged fluxes between 1 and 4~AU in the ecliptic plane,
the Harvard Radio Meteor Project's (HRMP) collection of meteor orbits, and
the zodiacal light measured from Earth and by Helios at a few locations
and in a few directions. The Divine model was eventually composed
of five populations of interplanetary meteoroids.

In the meteoroid model \citep{Staubach-1996,Staubach-et-al-1997,Gruen-et-al-1997},
data from the Galileo and Ulysses experiments were incorporated up to
Galileo's entry in the Jovian system, plus Ulysses' one year of data
centered at its first ecliptical plane crossing in 1995, obtained from
the ecliptic latitudes from~$-79^\circ$ to~$79^\circ$.
More information was retrieved from the Galileo and Ulysses data,
i.e.\ the direction, mass and speed of the impactors.
These two data sets only were used exlpicitly,
yet two components of the Divine model were imported,
the so-called ``core'' and ``asteroidal'' populations. The ``core''
population absorbed as much data as it was possible to fit with a single
mathematically-separable function~$H\times p_q\times p_e\times p_i$,
and since it slipped into the Staubach model, the latter inherited
partially the database of the Divine model. In the Staubach model,
a population of interstellar dust discovered earlier \citep{Gruen-et-al-1994}
was introduced as well.

The new meteoroid model is in no part based on the zodiacal light data
any more. Instead, the infrared observations by the COBE near-Earth
observatory are adopted as the new model base.
This replacement leads to improvement of the model
in two ways. First, the COBE maps of infrared sky provide a wide
spectral and surface coverage, while \cite{Divine-1993} used a handful ($<10$)
of the zodiacal light intensities picked up from the Earth-based and
Helios data sets, all measured at the same wavelength. Second,
the visual-wavelength emission from interplanetary meteoroids is
significantly more difficult to simulate than the thermal emission,
meaning more assumptions and more uncertainties in modeling the zodiacal
light data. For example, in the Mie theory for spherical particles,
the thermal emission is calculated based on the absorption
coefficient alone, while the visual light data simulation necessitates
the full phase curve. The results of the transition
to the IR data are more reliable, more in-depth simulations
of a quantitatively superior meteoroid data set.

The in-situ impact counts by the Galileo and Ulysses dust detectors
are incorporated in the new model. The Ulysses data are up to the end of 2003,
including the second near-perihelion ecliptic plane crossing. The data
are simulated in more detail than by Divine, taking the direction information
into account, yet with less assumptions than in the Staubach model
about the mass and speed of the impactors inferred through rather uncertain
relations from the raw detector measurements of impact-generated plasma.

The model of interplanetary flux at 1~AU by \cite{Gruen-et-al-1985} is incorporated
in the new model, too. The treatment of the model is different from that of the
previous works, however. The cumulative mass distribution of meteoroids in
the flux on spinning plate at 1~AU in the ecliptic plane
derived originally from the micro-crater counts on the lunar rock samples
delivered to Earth by the Apollo missions is converted back
into the raw crater size distributions. Then
the model is fitted to the raw distributions, taking both the mass and
speeds of meteoroids into account when predicting the crater sizes.

The orbital distributions of meteoroids inferred from the AMOR survey
of radio meteors was the key data set to replace the old HRMP data.
The reduction of radio meteor data involves correction for many
atmospheric biases, the most recent models of which were implemented
to obtain the true space distributions of meteoroids of the mass~$3\times10^{-7}$~g.
However, the result of this reduction suggested a cloud with
too many particles on highly inclined prograde orbits, in strong
contradiction to the latitudinal number density profile behind the COBE sky maps.
The model is therefore not based on the AMOR orbital distributions,
discussion of the radio meteors is nevertheless provided
in this paper.

\subsection{COBE/DIRBE Earth-bound infrared sky maps}

The COBE satellite was launched in orbit around the Earth with a set
of instruments on board to explore the cosmic background radiation~\citep{Bogges-et-al-1992}.
The DIRBE instrument, in particular, was included to search for the
cosmic {\em infrared} background. Observations of other sources of infrared emission
were second to this goal, so was the zodiacal dust cloud~\citep{Kelsall-et-al-1998}.

The DIRBE instrument took simultaneous observations through ten wavelength filters,
at 1.25, 2.2, 3.5, 4.9, 12, 25, 60, 100, 140 and 240 micrometers,
using a $0.7\times0.7$ square degrees field of view.
The radiation due to interplanetary dust is seen through all filters,
however, at the short wavelengths (1.25--3.5~micrometers), the infrared emission
is dominated by the point sources and diffuse sources other than
the interplanetary dust cloud. Moreover, the emission of interplanetary dust
includes both thermal and non-thermal components (i.e., light scattering) at these wavelength.
At the longest wavelengths (140 and 240 micrometers),
the infrared emission from the Milky Way and multiple diffuse sources
is paramount. The interplanetary dust thermal emission
is the main contributor between 4.9 and 100 micrometers only,
except at low galactic latitudes.

The DIRBE instrument was subject to special viewing constraints to prevent
its saturation by the bright sources such as the Sun and Moon.
In particular, it could never observe at the solar elongations
less than $60^\circ$ and greater than $130^\circ$.
Under this configuration, the emission from dust located
inside 0.86~AU from the Sun, the minimum distance seen from 1~AU
at the solar elongation~60$^\circ$, could not be detected.

The Earth's orbital plane is inclined to the plane of symmetry of zodiacal cloud
and the motion of Earth above and below the cloud's symmetry plane
affects the polar brightness by as much as 25\% \citep[Fig.~5 of][]{Kelsall-et-al-1998}.
The vertical motion can be exploited to measure quite precisely the inclination
($2^\circ$) and longitude of node ($78^\circ$) of the cloud's symmetry plane with respect
to the ecliptic. This was done in the framework of an empirical model by \cite{Kelsall-et-al-1998}
and their parameters are used to account for the vertical motion of the Earth,
when making predictions of the COBE observations.
The new model's symmetry plane is assumed in this case to match the zodiacal
cloud symmetry plane of the empirical model \citep{Kelsall-et-al-1998},
while in all other cases it is assumed to match the ecliptic plane---a
minor difference that can be neglected when predicting fluxes on spacecraft,
which is the model's ultimate goal.

In order to simplify the task of predicting the COBE/DIRBE observations in the meteoroid
model, we excluded from our considerations the wavelengths where the theoretical predictions
of the non-thermal radiation from dust are necessary, and where the interplanetary dust
is not the primary contributor, i.e., where dedicated effort to subtract emission
of the other objects properly would be necessary. Observations taken in five bands
at 4.9, 12, 25, 60 and 100 micrometers were thus selected for incorporation.

The COBE/DIRBE data come in variety of formats allowing for choice between low-level
intensities taken by individual exposures and extensively processed products such as an empirical
model of the zodiacal emission.
A compromise between the level of detail and amount of work
(the full size of the data set is measured in gigabytes) was found
in adopting weekly-averaged sky maps for further processing, yet
before any modeling of the emission sources is applied (zodiacal dust, etc.).

The thermal emission of interplanetary dust is predicted with help of
the Mie scattering theory. The dust particles are assumed to possess
spherical shape and to consist of ``astrosilicate''~\citep{Laor-Draine-1993}.
The material density of 2.5~g~cm$^{-3}$ is used when translating the grain masses into radii.
The temperatures and volume emissivities are calculated for the
heliocentric distances, particle radii and wavelengths of interest
using the thermal balance equation and the grey-body emission law,
taking the response function of the COBE/DIRBE filters into account.
We have tested our calculations successfully against the previously
accomplished works, specifically~\citep{Reach-1988},
and an unpublished manuscript by S.~D.~Price, P.~Noah and F.~O.~Clark
devoted to modelling the MSX observations~\citep{Price-et-al-2003}.
The volume emissivities are then convolved with the number density
of dust along the line of sight to predict the intensity measured
by DIRBE.

\subsection{Galileo and Ulysses DDS fluxes}\label{G&U DDS}

Galileo and Ulysses carried instruments for the direct dust impact detection
in the outer Solar system~\citep{Gruen-et-al-1997}.
The heavy Galileo spacecraft was launched into an initial orbit
demanding several gravitational maneuvers (at Venus and Earth)
in order to reach distant Jupiter. Each maneuver modified the orbit
of the spacecraft and its heliocentric velocity, thereby
adjusting detector sensitivity to dust particles on different orbits
and facilitating discrimination between them.

The light-weight Ulysses spacecraft was sent to Jupiter without
intermediate encounters with planets. The Jupiter fly-by resulted
in deflection of the spacecraft into a highly inclined orbit
allowing exploration of the dust cloud at high latitudes,
and at an unprecedented high relative velocity near the ecliptic plane.

The instruments on board the two spacecraft share the design of
the Dust Detector System, DDS~\citep{Gruen-et-al-1992b,Gruen-et-al-1992a}.
They had only minor differences in impact parameter
digitization procedures~\citep{Gruen-et-al-1995}.
The DDS measures the parameters of the plasma cloud released by
meteoroid impact, from which one can derive the masses and
impact speeds of the particles. At the impact speed of 20~km~s$^{-1}$,
the DDS can register meteoroids with the masses above~$2\cdot10^{-15}$~g.
At the same speed, meteoroids with the masses above~$2\cdot10^{-9}$~g
cause instrument saturation, so that only the lower limit of the mass
can be established. Such impacts are extremely infrequent onto
the 0.1~m$^2$ area of the target, however.

The Galileo and Ulysses data come in the same format.
The impacts for which detailed information was retrieved
(measured by the ion and electron sensors parameters of the plasma cloud
released by impact, spacecraft spin phase, and the mass and impact speed
determined from the plasma parameters) form tables of genuine impacts.
However, due to low transmission rates from
the Galileo spacecraft and during high-rate Jupiter dust streams
on both spacecraft not all the data about impacts were sent back to Earth.
The memory of DDS is limited and the data on old impacts
were overwritten when new impacts occurred before transmission.
On such occasions, the DDS built-in counters
only saved the total numbers of registered events categorized by
the quality class and measured ion charge amplitude.

Both Ulysses and Galileo are spin-stabilized spacecraft
with the spin axes being changed rather infrequently.
The two spacecraft rotate about the axes with the periods
orders of magnitude shorter than the typical time
between interplanetary and interstellar dust impacts.
It is reasonable to represent, therefore, the rotating dust detector
by a set of ``virtual'' detectors having fixed orientations
corresponding to different spin angles taking flux measurements
``simultaneously'' at the same location of spacecraft.
The exposure times of the virtual detectors are only fractions
of the exposure of the real detector though.

For those few occurrences when the impacts were reported by the counters only,
no virtual detector with fixed orientation is introduced
since some or all impacts can not be categorized by spin angle. A spin-averaged
dust detector sensitivity is then employed and the impacts are sorted in
time bins only. This is also done for the times when no impact was registered,
since it is then superfluous to make model predictions
for the virtual detectors all of which
would have gotten zero impacts on the data side.

The DDS instruments measure the mass and speed of meteoroids
indirectly, by determining them from the properties of the plasma cloud released
by impact. This determination is based on the ground calibration of instrument
and is prone to large uncertainties. The speed of meteoroids is inferred
from the rise times of plasma signals, and is uncertain by a factor of 2.
The mass is derived from the full charge released by impact,
after reduction for the poorly determined impact speed.
The uncertainty of mass determination is thus larger,
reaching one order of magnitude due to random errors of measurements
and one more order due to the systematic errors of particle material uncertainty.

Because of these uncertainties the mass and speed distributions of
meteoroids inferred from the raw detector data are not used to adjust
the new meteoroid model. The model is used instead to predict fluxes
of meteoroids that release plasma clouds larger than
a certain threshold charge magnitude, and all impacts with magnitudes
above that threshold are selected from the data, correspondingly.
According to the ground calibration, the ion charge~$Q_{\rm I}$ measured
by the dust detector is related to the impactor mass~$m$ and speed~$v$
via~\citep[a fit to the calibration plot in Fig.~3 of][]{Gruen-et-al-1995}
\begin{equation}
   \left( {Q_{\rm I} \over 1 \; \mbox{C}} \right) \approx 6\cdot10^{-5}
   \left( {m   \over 1 \; \mbox{g}} \right) \cdot
   \left( {v   \over 1 \; \mbox{km/s}} \right)^{3.5}.
\end{equation}

The bottom limit of measurable ion charge~$\mbox{min}\; Q_{\rm I}=10^{-14}$~C.
At 20~km~s$^{-1}$ impact speed, this corresponds to the minimum
meteoroid mass~$5\cdot10^{-15}$~g, or $0.1\mu$m. This is rather low for a
meteoroid model based on the Poynting-Robertson approximation for $M<10^{-5}$~g
\citep{Wilck-Mann-1996,Wehry-Mann-1999}.
In order to make sure that the dynamics of the impactors
are compatible with those of the model meteoroids, the weakest impacts
are not taken into account. The DDS categorizes all impacts by the ion charge
amplitude into six amplitude ranges AR from 1 to 6. We use the bottom limit
of the AR3, that the ground calibration sets at $Q_{\rm I}=1.12\cdot10^{-12}$~C for Galileo
and $Q_{\rm I}=1.03\cdot10^{-12}$~C for Ulysses.
Note this implies that only the bigger impacts are taken into
account, unlike the model by \cite{Staubach-1996}.

Since after the development of the Staubach model it was recognized that the
wall impacts into DDS produce events that are indistinguishable from the
target impacts~\citep{Altobelli-et-al-2004}, especially for the bigger impactors,
the DDS on Galileo and Ulysses was approximated by a flat plate of 0.1~m$^2$ area rather than
by the wall-shielded target.

The ground calibration of DDS could be performed with the small impactors only.
Thus the thresholds specified above were extrapolated from the laboratory measurements
taken with the small grains at higher speeds, rather than with the big grains
at lower speeds. When the DDS data from both spacecraft were incorporated
in the model simultaneously with the COBE infrared sky maps,
they turned out to be incompatible. Assuming that the COBE maps
were modeled adequately and that the extrapolation of ground calibration of
DDS was valid, we would obtain fluxes 2.5 higher than those actually reported.
When favouring the DDS data and calibration, we would get a correspondingly
darker infrared sky.

The decision was made to revise the ion charge $Q_{\rm I}$ necessary to generate AR$\geq3$ events
upward by a factor of~16, that is to require an impactor 16 times more massive than it was
believed before, or 2.5 greater in size. The revision was sufficient to lower the flux
of dust grains in the Poynting-Robertson regime and to eliminate the discrepancy between
DDS and COBE data.

Similar revisions were suggested by the comparison of DDS and Pioneer~10
and~11 data (M.\ Landgraf, private communication) and by the determination
of particle sizes in the Jovian dust rings based on optical and in-situ
measurements from Galileo (H.\ Kr\"uger, private communication),
although no affirmative conclusion was drawn yet in either case.

We note that in our case the revision is forced by the cumulative mass distribution
taken from the interplanetary meteoroid flux model~\citep{Gruen-et-al-1985}
that strictly binds the number of particles detected by DDS and the cross-section
area contributing to COBE observations.

\subsection{The lunar micro-craters}

The interplanetary meteoroid flux model~\citep{Gruen-et-al-1985} specifies the cumulative
mass distribution of particles in the flux onto a flat plate orbiting the Sun at the Earth's
distance in the ecliptic plane and spinning about the axis perpendicular to that plane (see the ``total''
flux curve in Fig.~\ref{Mass Distrib}).
The model is based on the size distribution of micro-craters found on the lunar rock samples
returned by the Apollo missions, plus several early spacecraft dust detector results.
The conversion of the crater diameter~$D$ into the impactor size was performed using relationship
\begin{equation}
   D = cm^\lambda,
\end{equation}
where $c=8.24$ and $\lambda=0.37$ for an impact speed $v=20$~km~s$^{-1}$.

In order to allow for more flexibility of model distributions, especially velocity distribution
of meteoroids, the interplanetary meteoroid flux model was converted back into the
crater size frequencies, and the frequencies were used as a target for model fit.
The impact speeds of meteoroids were taken into account when making model predictions
of the crater frequencies~\citep{McDonnel-1978}
\begin{equation}
   {D_1 \over D_2} = \left( {v_1 \over v_2} \right)^{0.67}.
\end{equation}

\subsection{The AMOR radio meteor survey}

The AMOR radar~\citep{Baggaley-2001} measures echoes from meteors ablating
in the Earth atmosphere and from these measurements the elements
of pre-encounter heliocentric orbit are
inferred. All registered meteors are archived in a database. The database can be
used to produce the distributions of observed meteors in orbital elements.
The correspondence between the observed distributions and the true space distributions
at a mean mass threshold of $\sim3\cdot10^{-7}$~g (20~$\mu$m~radius)
is established through a sophisticated bias correction procedure~\citep{Galligan-Baggaley-2004}.
The corrected distributions were produced in the form
of three-dimensional array of numbers of meteoroids in rectangular cells
covering semimajor axes from 0 to 6~AU, eccentricities from 0 to 1,
and inclinations from 0 to 180$^\circ$,
in Earth-crossing orbits only.

It was realised, however, that the AMOR corrected distributions
contradict to the COBE/DIRBE data and that it is not possible
to fit the two data sets simultaneously under the model assumptions.

\begin{figure}[h]
\centerline{\psfig{figure=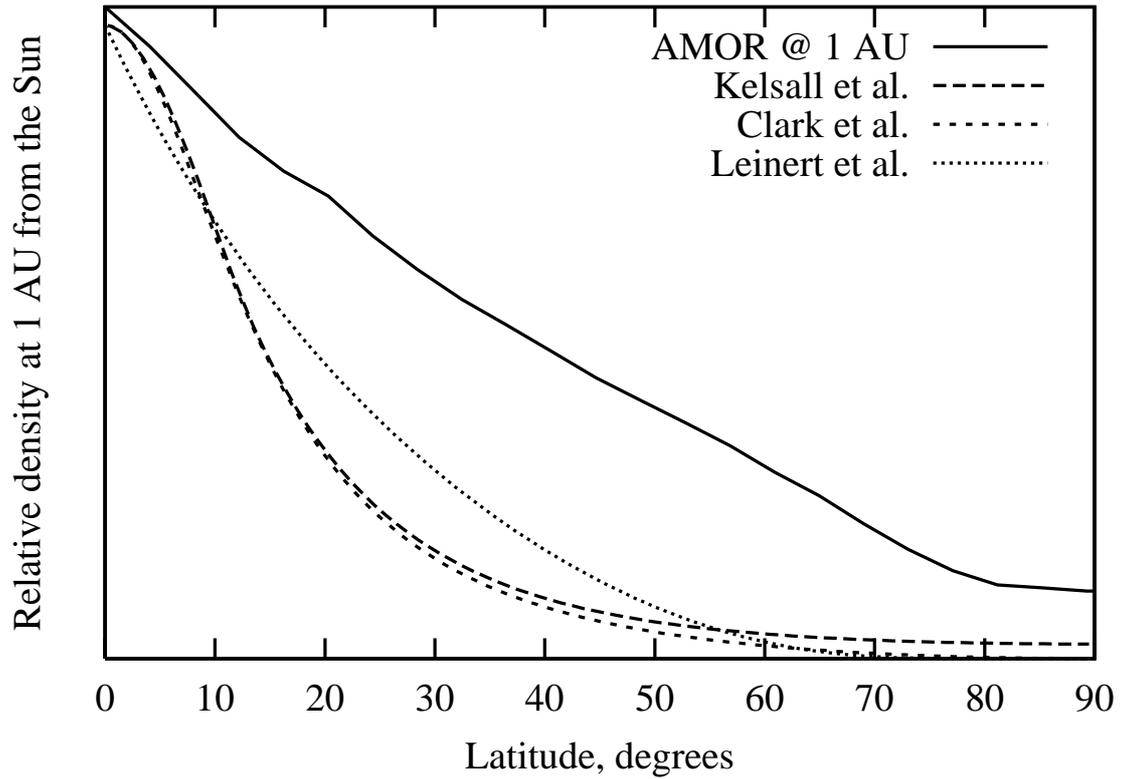,width=.7\hsize,angle=270}}
\caption{The latitudinal number density profiles inferred from the
infrared observations by IRAS~\citep{Clark-et-al-1993},
COBE~\citep{Kelsall-et-al-1998}, zodiacal light measurements~\citep{Leinert-et-al-1981},
and built based on the AMOR corrected orbital distributions~\citep{Galligan-Baggaley-2004}
at the distance 1~AU from the Sun.
\label{latprofiles}}
\end{figure}

Fig.~\ref{latprofiles} compares the latitudinal number density profiles
of the zodiacal cloud at 1~AU obtained from different data sets by different
teams. The distance of 1~AU from the Sun was chosen as the only one
where the direct conversion of the AMOR orbital distributions into
the number density is possible without extrapolations. Note, however,
that in the alternative models~\citep{Leinert-et-al-1981,Clark-et-al-1993,
Kelsall-et-al-1998} the number density is a-priori mathematically separable
in radial distance and latitude.

%
%

The plot makes it obvious that the AMOR distributions determine too wide a
number density profile, incompatible with the other data sets, including COBE.
One might argue that the radar is sensitive to the meteoroids of sizes
different from the major contributors to the infrared emission and zodiacal
light. In the new meteoroid model, however, both the radar meteors and
the infrared emission are dominated by the particles of masses $<10^{-5}$~g,
which is a consequence of using the mass distribution of~\citep{Gruen-et-al-1985}.
This mass distribution ensures that 80\% of the total particle cross-section area
(which the zodiacal light and infrared emission should be proportional to)
belongs to the meteoroids from 10 to 100~$\mu$m in radius
($10^{-8}$ to $10^{-5}$~g), the range embracing the meteoroid sizes
to which the AMOR is sensitive, and adequately described in the meteoroid model
in the Poynting-Robertson regime.

Until the discrepancy between the AMOR orbital distributions and
COBE latitudinal profile is resolved, either by improvements
of the AMOR corrections or further development of the model
populations, the AMOR distributions can not be incorporated in the new model.
We note that that out-of-ecliptic broad pattern is found in
many previous radar surveys~\citep[see][and references therein]{Steel-1996}.

The reason of the discrepancy may also be in the material density and albedos
of the particles adopted here, as well as in their constancy assumption.

\section{The best-fit model}

\subsection{The Gaussian likelihood estimator}

The observations of electromagnetic emission from the zodiacal cloud particles
are based on registration of photons. The number of the photons contributing
to a single measurement as well as the number of dust particles that have
emitted them are very large. The observation noise in this case
is best described by Gaussian statistics.

The Gaussian distribution provides the probability density to obtain
the observation~$O_j$ given its expectation~$C_j$ predicted by model
($j$~is the observation index, $j = 1,\ldots,M$)
\begin{equation}
   P_{\rm Gauss} (O_j/C_j) = {1\over\sqrt{2\pi\sigma_j}} \exp
   \left\{ - {(O_j-C_j)^2 \over 2\sigma_j^2} \right\}
\end{equation}
where $\sigma_j$ is the standard deviation of the $j$-th observation.

The model prediction~$C_j$ is, in fact, a sum of contributions
due to $N$ meteoroid populations
\begin{equation}
   C_j(\vec x) = \sum_{i=1}^N A_{i,j} x_i
\end{equation}
where $A_{i,j}$ is the $i$-th population's share in
the $j$-th observation before the normalization
and $x_i$ is the $i$-th population's weight.

For a set of independent observations $O_j$, $j = 1 \ldots M$,
the likelihood estimator
\begin{equation}
   \Pi_{j=1}^M P_{\rm Gauss} (O_j/C_j)
\end{equation}
is typically introduced.

The logarithm of likelihood function looks more simple
and is actually used as estimator
\begin{equation}\label{Gaussian likelihood}
   G (\vec x) = {\rm ln}\; \Pi_{j=1}^M P_{\rm Gauss} (O_j/C_j(\vec x)) =
   - \sum_{j=1}^M {\left( \sum_{i=1}^N A_{i,j} x_i - O_j \right)^2 \over 2\sigma_j^2}
\end{equation}
where the terms ${\rm ln} \; \sqrt{2\pi\sigma_j}$ are omitted because
they do not depend on $\vec x$.
The gradient of likelihood criterion~$G$ is given by
\begin{equation}\label{Gaussian gradient}
   {\partial G \over \partial x_k} = - \sum_{j=1}^M {A_{k,j}
   \left( \sum_{i=1}^N A_{i,j} x_i - O_j \right) \over \sigma_j^2}.
\end{equation}

The infrared observations taken by the COBE satellite are provided with
estimates of accuracy in the form of standard deviations. These deviations
are typically very small, well below the systematic errors of
the meteoroid model predictions. Higher standard deviations~$\sigma_j$
are therefore introduced artificially to bracket a tolerable range
of systematic model deviations, and are set to 10\% of the maximum
brightness for each filter.

\subsection{The Poissonian likelihood estimator}

Meteoroid counts with a dust detector or meteor radar are subject
to different kind of noise because the measurements result in
small integer numbers. The Poissonian distribution is applicable in this case.
It gives the probability to count~$O_j$ events given the theoretical
prediction of~$C_j$ events
\begin{equation}
   P_{\rm Poisson} (O_j, C_j) = {C_j^{O_j} \over O_j!}
   \exp \left\{ - C_j \right\}.
\end{equation}

The likelihood criterion of model predictions~$C_j$ is defined as
in the previous case, its logarithm is
\begin{equation}\label{Poissonian likelihood}
   P (\vec x) = {\rm ln}\; \Pi_{j=1}^M P_{\rm Poisson} (O_j, C_j(\vec x)) =
   \sum_{j=1}^M \left[
   O_j {\rm ln}\; \sum_{i=1}^N A_{i,j} x_i - \sum_{i=1}^N A_{i,j} x_i
   \right]
\end{equation}
where the constant terms ${\rm ln}\; O_j!$ are omitted.
The gradient of likelihood criterion~$P$ is given by
\begin{equation}\label{Poissonian gradient}
   {\partial P \over \partial x_k} = \sum_{j=1}^M \left[
   {O_j A_{k,j} \over \sum_{i=1}^N A_{i,j} x_i} - A_{k,j} \right].
\end{equation}

\subsection{The likelihood estimator for heterogeneous data}

The meteoroid data sets are composed of both electromagnetic
emission observations with the Gaussian distribution of errors
and individual particle counts with the Poissonian distribution.
In order to incorporate all data simultaneously,
a composite likelihood function is introduced in the form of weighted
sum of the likelihood functions for individual data sets.
The weights are used to emphasize certain data sets that would
otherwise contribute too weak to the composite likelihood function.

\subsection{The best fit}

The COBE/DIRBE infrared sky maps, Galileo and Ulysses DDS impact counts, and the
interplanetary meteoroid flux model were fed into
the inverse problem solution program. Different weights were assigned to
each data set.

The COBE data were assigned weight~1.
The number of pixels in each of the five single-wavelength maps incorporated
was about $2\cdot10^4$.
The Galileo and Ulysses data were not influential in such environment,
and their weights were raised to~50.
The lunar micro-crater counts were set to 1000~craters
above each threshold, giving these measurements already
a low standard deviation under the Poissonian hypothesis,
and their likelhood function was additionally multiplied by~20.

The composite likelihood estimator function was maximized using the L-BFGS-B library routines
\citep{Zhu-et-al-1997} implementing a sophisticated branch of
the Powell (gradient descent) algorithm with constraints.
The normalization factors were not allowed to decrease below zero.
The optimal factors were found and the original (non-normalized)
populations were scaled accordingly. The resulting set of weighted
populations represented the best-guess meteoroid model.

Since the best fit found this way was not universally realistic,
showing rather strong variations of the normalisation factors
of the adjacent populations of dust from comets, a smoothing
criterion was introduced to complement the likelihood estimator
for heterogeneous data. The additional criterion was stated as
\begin{equation}
   S(\vec x) = \alpha \sum_{i=1}^N \sum_{j=1}^N \Delta_{i,j}
   \left( {\rm ln}\; [\varepsilon+x_i] - {\rm ln}\; [\varepsilon+x_j] \right)^2,
\end{equation}
where $\Delta_{i,j}=1$ when the populations $i$ and $j$ occupy
the adjacent ranges of encounter speed with Jupiter~$U$,
and zero otherwise. A small value of $\varepsilon=10^{-10}$ was added
to the population weights in order to prevent underflows when
$x_i=0$. The parameter $\alpha=10$ was found
experimentally so as to minimize its impact on the quality of fit,
yet reach a realistic smoothness.

\begin{figure}[h]
\centerline{\psfig{figure=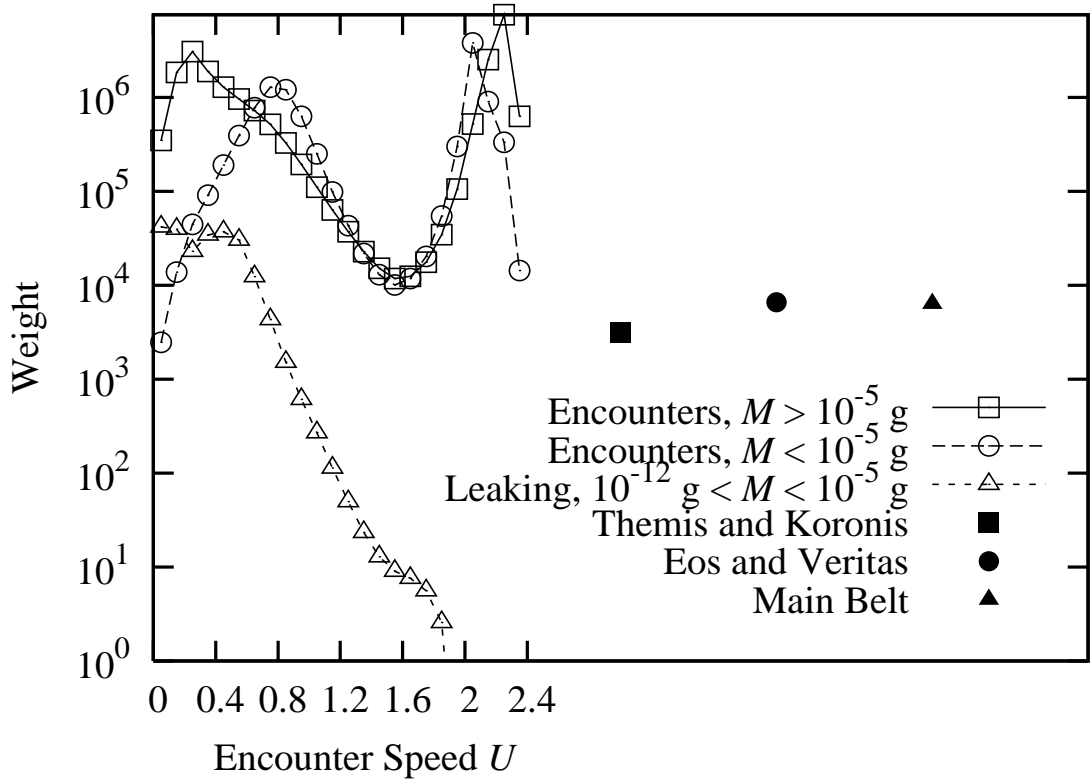,width=.7\hsize,angle=270}}
\caption{Parameters of the final fit of the new meteoroid model. The three filled
symbols show the weights of the three populations of dust from asteroids, spiraling
under the Poynting-Robertson effect. The collisionally evolving populations are not
shown since their weights are bound in the model
to the corresponding Poynting-Robertson dust populations.
The empty symbols depict the weights of the populations of dust from comets,
both in the regime of encounters with Jupiter (with the collisional
and Poynting-Robertson mass distributions above and below $M=10^{-5}$~g)
and leaking from the region of encounters toward the Sun.}
\label{bestfit}
\end{figure}

The final fit parameters are displayed in Fig.~\ref{bestfit} and refer
to the orbital distributions introduced in Sect.~3 and normalised so that
their integrals before the fit over the orbital space are unit.
The number density of the interstellar dust particles greater than $10^{-14}$~g
in mass was found to be $2.8\cdot10^{-9}$~m$^{-3}$, very close to
the implications of the flux inferred
from the Ulysses measurements by \cite{Landgraf-et-al-2000}.
According to our fit, the populations of dust from asteroids are mediocre
in the environment dominated by the dust from comets. The populations of
dust from comets belonging to different categories (leaking-encoountering,
big-small) show remarkable correlations of their~$n(U)$ profiles,
despite no relevance was supposed a-priori.

\begin{figure}[h]
\centerline{\psfig{figure=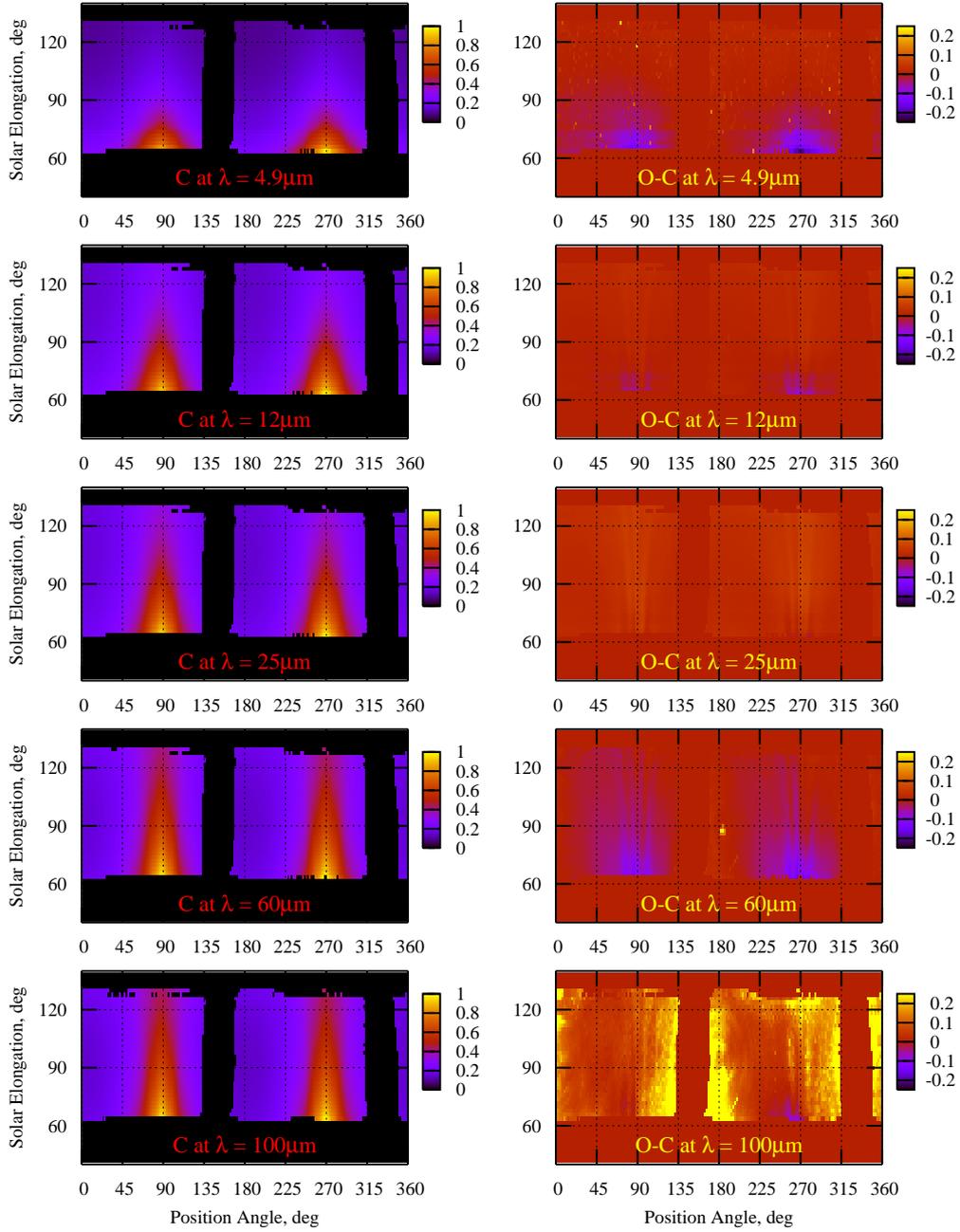,width=0.9\hsize}}
\caption{The best-fit model of the five COBE/DIRBE infrared sky maps (left), and its residuals,
observations minus calculations (right), in relative units. Black areas on the left pannels
are the forbidden zones of high and low solar elongation
and the galactic plane excluded from the data set to avoid interference.}
\label{cobe}\label{fit-first}
\end{figure}

\begin{figure}[h]
\centerline{\psfig{figure=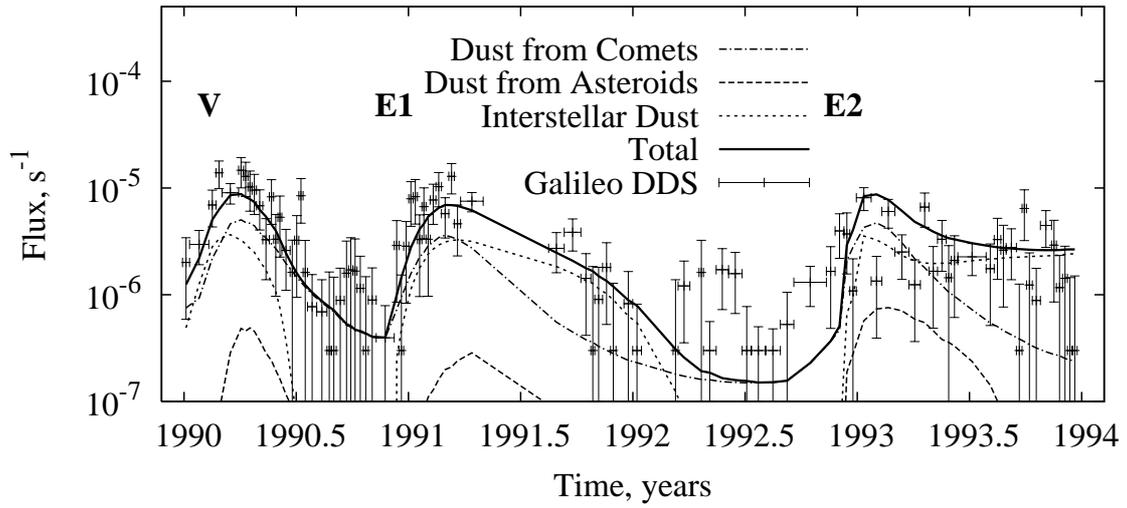,width=0.45\hsize,angle=270}}
\caption{Impact rates measured by the Galileo dust detector, spin-averaged. Labels
`V', `E1' and `E2' mark one Venus and two Earth gravitational
maneuvers, respectively.\label{galileo}}
\vspace{5cm}
\end{figure}

\begin{figure}[h]
\centerline{\psfig{figure=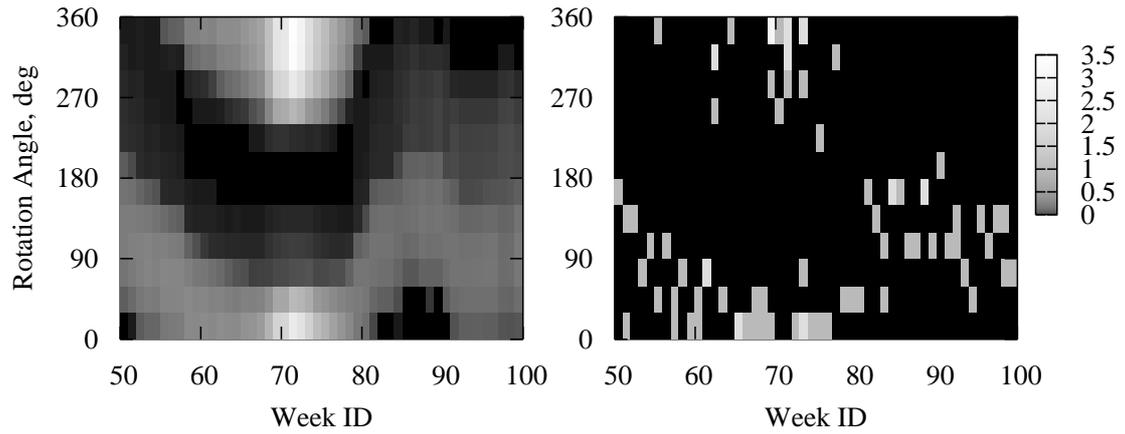,width=0.38\hsize,angle=270}}
\caption{Impact counts with the Ulysses dust detector. Model expectations
are on the left panel, actual counts are on the right panel. The time
frame is chosen so that the first near-perihelion ecliptic plane
crossing occurs on week~72.\label{ulysses}\label{fit-last}}
\vspace{5cm}
\end{figure}

%
%
The quality of the final fit can be assessed in Figures~\ref{fit-first}--\ref{fit-last}.
The comparison with COBE observations is shown in Fig.~\ref{cobe}. The infrared sky maps
are constructed versus solar elongation and ``position'' angle
with respect to the Sun. The latter is measured from the ecliptic
north clockwise about the Earth-Sun axis.
At the solar elongation of~$90^\circ$, the position angle of~$90^\circ$
is close to the apex of Earth motion,
the position angle of~$180^\circ$ is close to the vertex.

The agreement between data and theory is very good. For most pixels, the residual is below 10\%.
Some point sources not accounted by the interplanetary dust model are recognizable in the residual
maps at the wavelengths 4.9 and 60~$\mu$m, while the map at 100~$\mu$m is polluted by the Milky Way
that is still intensive near the band excluded along the Galaxy plane.

The spin-averaged impact rates on the Galileo dust detector~(Fig.~\ref{galileo})
are for the most time bins within the $1\sigma$ interval. The rates are dominated by
the dust from comets and interstellar space, while the asteroids
contribute less than 10\% of impact events.

%
%

The impact counts with the dust detector of Ulysses are shown
in Fig.~\ref{ulysses}. The impacts of interstellar dust grains occupy
the range of rotation angles from 0 to $180^\circ$
and are seen as a zig-zag-shaped grey area in the lower part of the maps.
Interplanetary dust impacts are dominant during the first
near-perihelion ecliptic plane crossing that happened on week~72,
the flux peak is at approximately $300^\circ$.

%
%

\subsection{The model test against the independent data}

The impact events collected by the dust detectors on board Helios~1
and Pioneer~11 satellites were of low statistical significance and
have not been incorporated in the new meteoroid model. However,
they can serve as an independent test of the model.

The Helios~1 spacecraft was launched into a highly eccentric orbit
about the Sun with the aphelion at 1~AU and the perihelion at 0.3~AU.
The orbital plane lay in the ecliptic plane \citep{Gruen-1981}.
The spacecraft spinned about an axis perpendicular to that plane.
Helios~1 made ten revolutions around the Sun over five years of operation,
with two dust instruments registering dust impacts coming
from the ecliptical plane and from the ecliptic south.
The dust instruments measured, among other characteristics,
the ion charge in the plasma cloud released by impacts. Although the
detailed information on released charges has not survived the time,
every impact in the records still available is marked
by the ion charge range to which the charge belonged.
The ion charge was related to the mass~$m$ and speed~$v$
of the impactor via the formula~\citep{Gruen-1981}
\begin{equation}
   \left( {Q_{\rm I} \over 1 \; \mbox{C}} \right) \approx 4.07\cdot10^{-5}
   \left( {m   \over 1 \; \mbox{g}} \right) \cdot
   \left( {v   \over 1 \; \mbox{km/s}} \right)^{2.7}.
\end{equation}

\begin{figure}[h]
\centerline{\psfig{figure=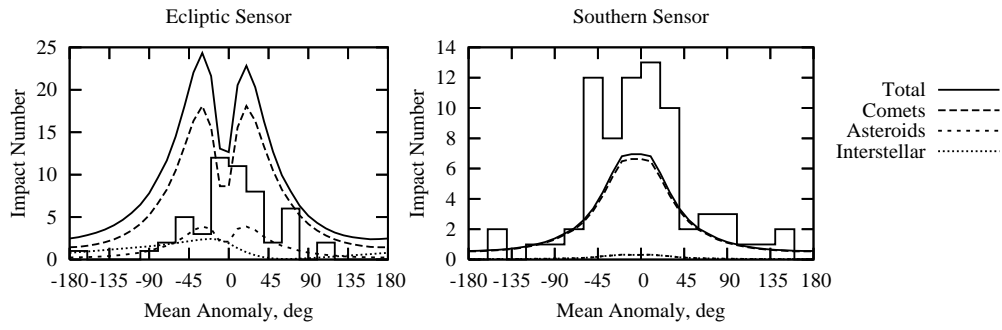,width=0.9\hsize}}
\caption{Impact counts with the dust detectors on board Helios~1.\label{helios}}
\vspace{10cm}
\end{figure}

We selected the impacts that fell in the ion charge ranges $\geq2$ ($Q_I\geq10^{-13}$~C) since
the lowest range~1 appeared to be dominated by very small grains not described
by the model quite well. The model expectations corrected for the ``black-out''
times ($\approx$39\%) when the data were not transmitted from satellite,
and the actual impact counts are shown in Fig.~\ref{helios}.
The theoretical prediction remains within a factor of two to three of the measurements.
The model predicts also a distinct feature of the local minimum of flux
near the perihelion of the orbit of spacecraft that is absent from the data,
although data statistics are rather poor. The local minimum is the consequence
of the decrease of the average velocity of meteoroids relative to spacecraft
causing the flux to fall as well. There is no such a minimum
of the model predictions for the southern sensor.

When making predictions for the ecliptic sensor, the effect of a thin foil
in its front was not taken into account, a necessary simplification
that may explain why the model expectations are higher, and why the local minimum
of flux is missing in the data while in the theory
two maxima are present around zero mean anomaly.
The ecliptic sensor was protected by the foil from the direct solar radiation,
but the foil was also causing a bias by preferentially deflecting small and oblique
impactors~\citep{Pailer-Gruen-1980}.

%
%

The Pioneer~11 spacecraft was sent first to Jupiter, where it made
a gravitational maneuver that deflected it into the path towards
Saturn. Pioneer~11 carried penetration dust detector sensitive to relatively
big meteoroids, $6\cdot10^{-9}$~g at 20~km/s~\citep{Humes-1980}.
The dust detector foil was punctured by the meteoroids satisfying
the inequality
\begin{equation}
   \left( {m   \over 1 \; \mbox{g}} \right) \cdot
   \left( {v   \over 1 \; \mbox{km/s}} \right)^{2.5} > 10^{-5}.
\end{equation}

\begin{figure}[h]
\centerline{\psfig{figure=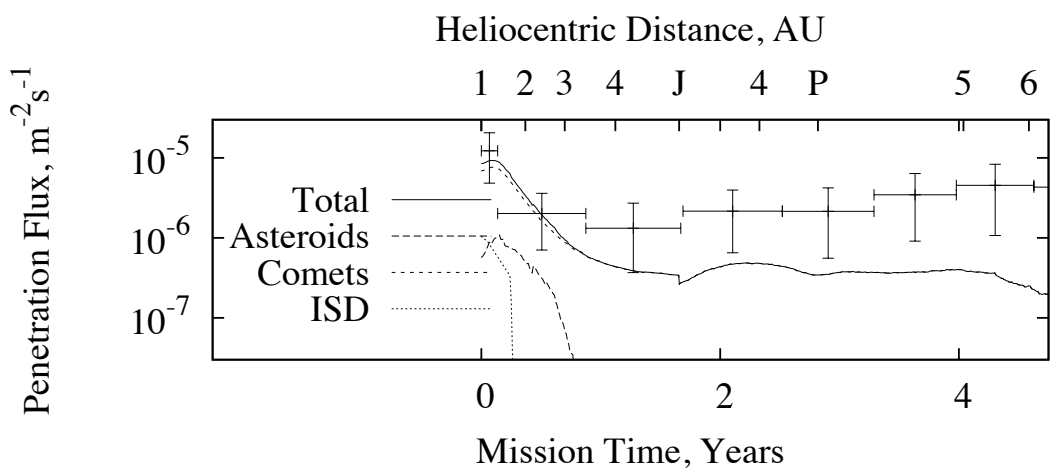,width=.93\hsize}}
\caption{Fluxes inferred from the Pioneer~11 impact counts,
taking the uncertainty of the number of active cells of dust detector
into account \citep{Dikarev-Gruen-2002} shown with error bars,
and the meteoroid model expectation plotted as a continuous curve. On the heliocentric distance
axis, `J' marks the Jupiter fly-by, and `P' marks the perihelion
of the post-Jupiter orbit.\label{p11}}
\end{figure}

The most intriguing result of this experiment was a flat penetration flux
measured ever since
the spacecraft left the asteroid belt behind. Although somewhat below
the measured value, the flat rate is predicted
by the new model too~(Fig.~\ref{p11}). According to the meteoroid model,
most of the impacts were due to the particles from comets, while the
dust from asteroids and interstellar space contributed each less than
10\% of observed fluxes.
A discrepancy between the model prediction and the data seen in Fig.~\ref{p11}
for heliocentric distances larger than 5~AU can be explained by a dominant contribution
from meteoroid sources beyond Jupiter \citep{Landgraf-et-al-2002}.

%
%

\subsection{Comparison with previous meteoroid models}

%
%

The new meteoroid model differs essentially from the previous tools to predict particle
fluxes on spacecraft in the way the orbital distributions of meteoroids are established.
While Divine and Staubach used the observational data only to constrain the distributions,
the new model comes with a wide set of meteoroid populations defined a-priori for
a number of possible sources, evolving under the assumed dynamical laws. For instance,
one name of meteoroid population, ``asteroidal'', in the Divine model was referring
solely to the proximity of the meteoroids to the asteroid belt, while in the new model
the populations of dust from asteroids have genetic relationship
to the main belt and prominent families.

\begin{figure}[h]
\psfig{figure=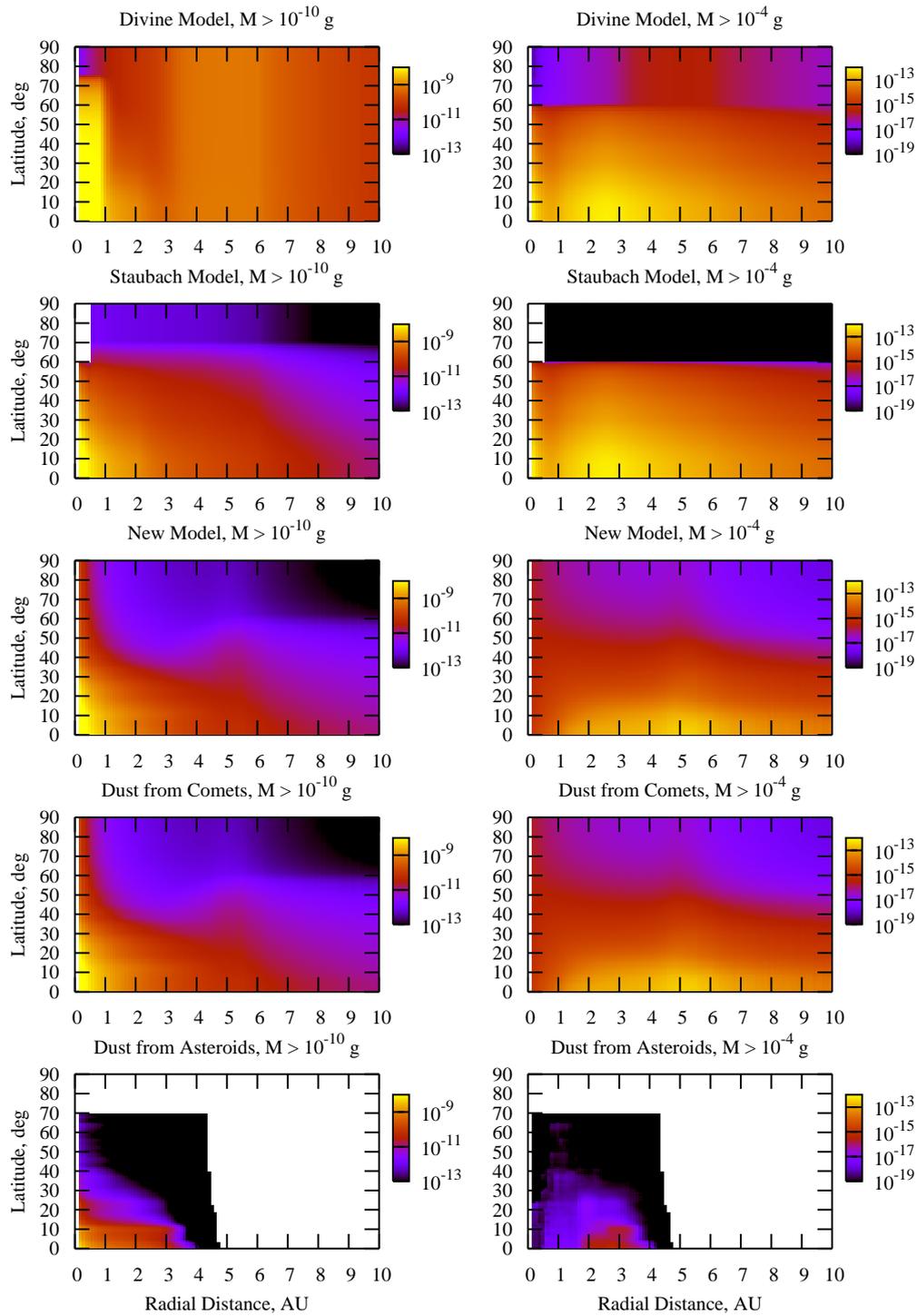,width=0.9\hsize}
\caption{The number density (m$^{-3}$) of meteoroids above two mass thresholds,
$10^{-10}$~g (left) and $10^{-4}$~g (right), predicted for the Divine,
Staubach and the new models. For the new model, shown are also
the number densities of meteoroids shed by comets and asteroids, respectively.
\label{snapshot}}
\end{figure}

The evolution of models can be viewed in Fig.~\ref{snapshot} where the number densities
of meteoroids are mapped for wide ranges of heliocentric distances and latitudes above the
ecliptic plane. Note that the number density at 1~AU in the ecliptic plane
are similarly predicted by all models for the small grains distributed by the Poynting-Robertson
effect. However, the new model specifies a substantially
lower density at that location of the big grains controlled by gravity.
In the new model, the majority of big particles are in Jupiter-crossing orbits,
and their velocities are high with respect to the Earth and Moon. Therefore,
in order to reproduce the lunar micro-crater frequencies, the new model needs
a lower number density of meteoroids.

In the Divine model, a set of rectangular number density ``patches'' is most
obvious, each of which corresponds to a single population having mathematically-separable
distributions in perihelion distance, eccentricity and inclination, and therefore separable
radial and latitudinal number density profiles. In the Staubach model predictions,
manifestation of separability is mitigated by the logarithmic color scale, like
in the Divine model predictions for the bigger meteoroids
(logarithm of separable function is not necessarily separable).

In the new model, non-separability is imminent. Since the comets were represented by
those on Jupiter-crossing orbits only, the number density of meteoroids peaks along
the orbit of the giant planet. At the same heliocentric distance, the latitudinal
number density profile is also widened. It looses breadth away from the Sun and down
to the asteroid belt toward the Sun, from where it is extending in latitudes again
(the effect is magnified greatly by the logarithmic scale but it is still real).

The asteroids are not separable as a source, too (see the plot for $M>10^{-4}$~g),
so is the dust spiraling from them due to the Poynting-Robertson effect,
until it leaves the source region. The ten-degree band due to the Eos and/or Veritas family
is best revealed in the map of dust from asteroids, so it contributes a little density to
the overall map as well.

\begin{figure}[h]
\psfig{figure=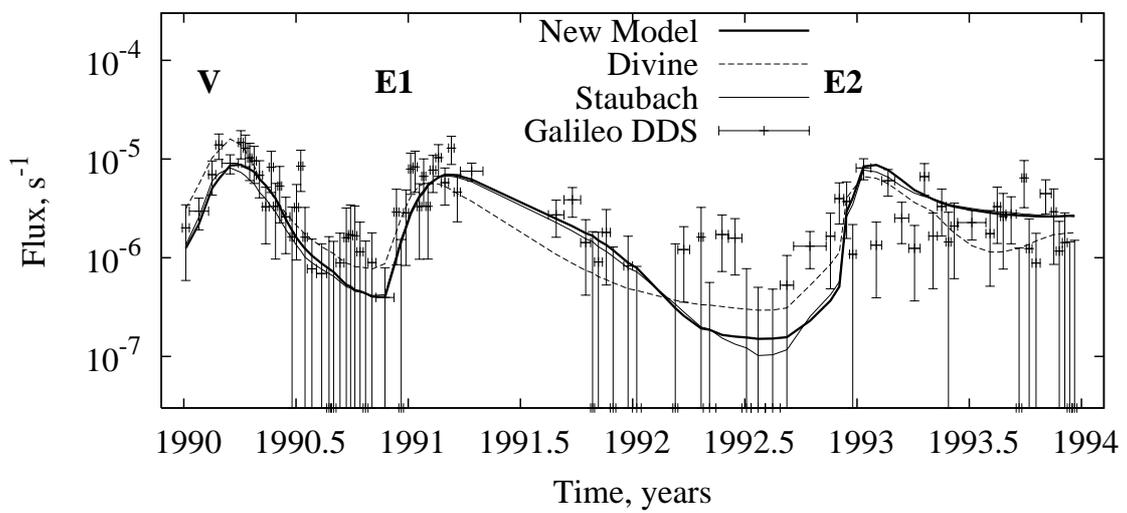,width=0.45\hsize,angle=270}
\caption{Impact rates measured by the Galileo dust detector, spin-averaged,
and their theoretical counterparts calculated using the Divine, Staubach, and
the new meteoroid models. Labels `V', `E1' and `E2' mark one Venus and two Earth gravitational
maneuvers, respectively.\label{galileo-models}\label{models-first}}
\end{figure}

\begin{figure}[h]
\centerline{\psfig{figure=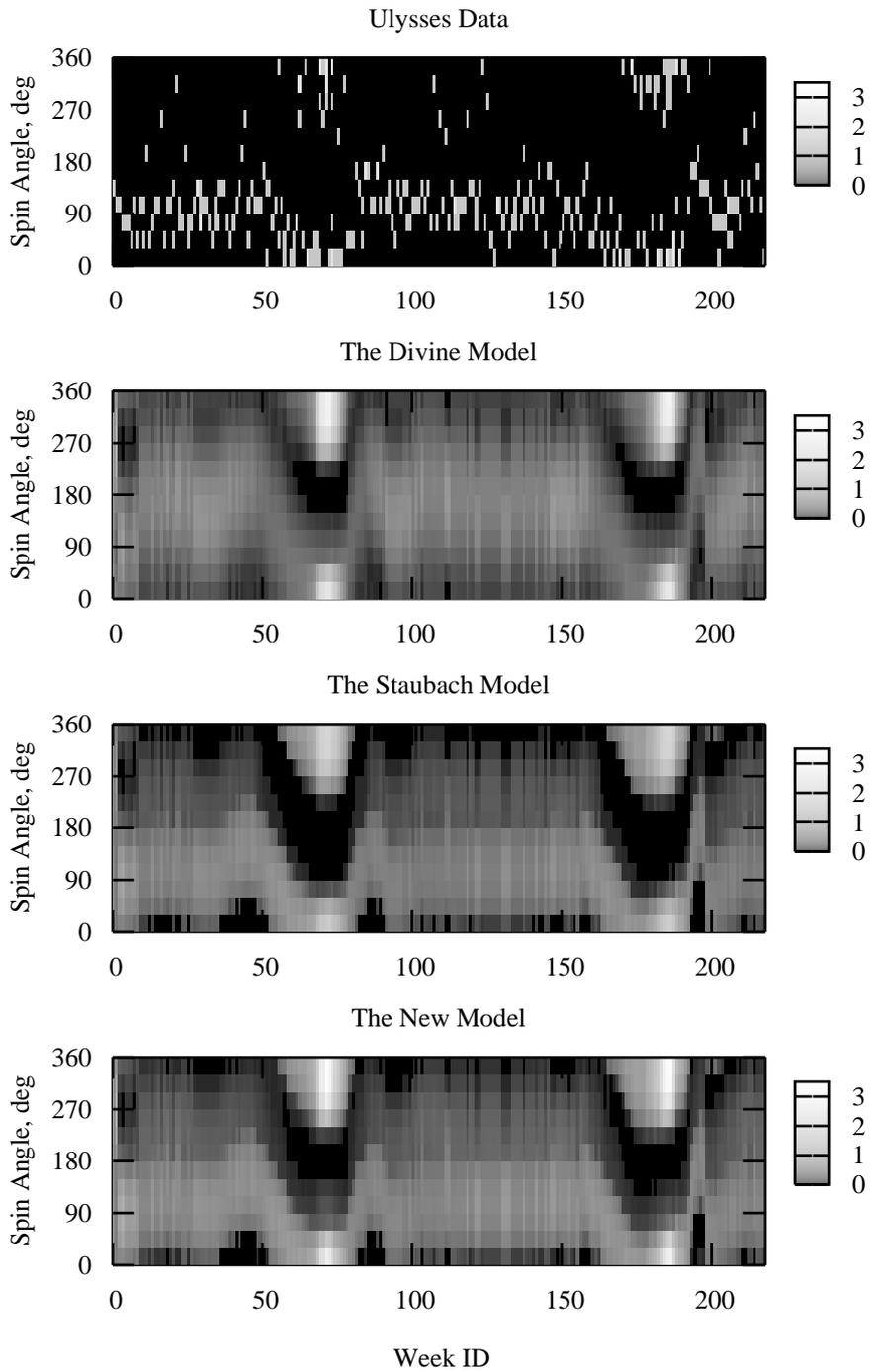,width=1.25\hsize,angle=270}}
\caption{Impact counts reported by the Ulysses dust detector, for those weeks
when at least one impact was detected, and their theoretical counterparts
calculated using the Divine, Staubach, and the new models. The gray levels
indicate the impact numbers.
\label{ulysses-models}}
\vspace{5cm}
\end{figure}

By no surprise, the new meteoroid model is good or better at representation
of those data sets incorporated. Interestingly, the Galileo DDS~(Fig.~\ref{galileo-models})
data are still reproduced by the Divine and Staubach models almost as good as by the new model,
in spite of our revision of the DDS sensitivity threshold. The Divine model lacks
the interstellar dust population that is obvious in the Ulysses data, however (Fig.~\ref{ulysses-models}).
The information on spin angles of the Ulysses impacts not incorporated in the Divine model
rejects Divine's ``halo'' population that is dominant away from the
ecliptic plane, e.g.\ week IDs 100--150. It is clearly seen that only the Staubach and new models
give the right spin angle distributions of impacts during these periods.

%
%

\begin{figure}[h]
\psfig{figure=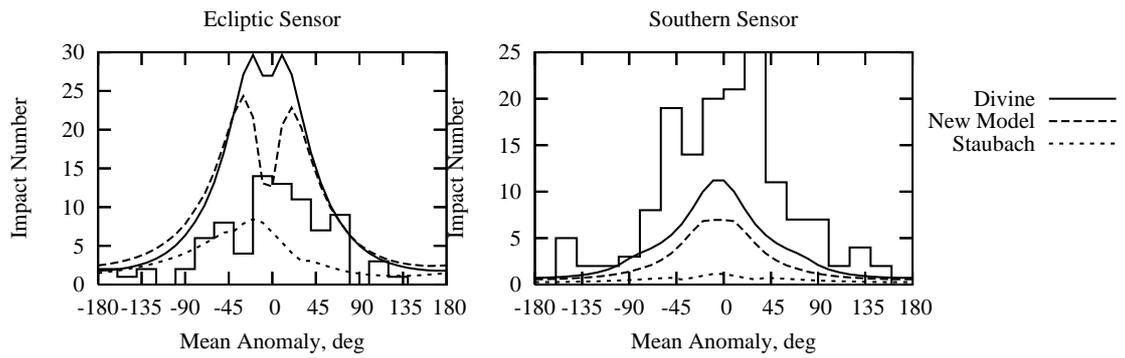,width=\hsize}
\caption{Impact counts with the dust detectors on board Helios~1 and their theoretical
counterparts calculated with the Divine, Staubach, and the new model.\label{helios-models}}
\vspace{5cm}
\end{figure}

\begin{figure}[h]
\centerline{\psfig{figure=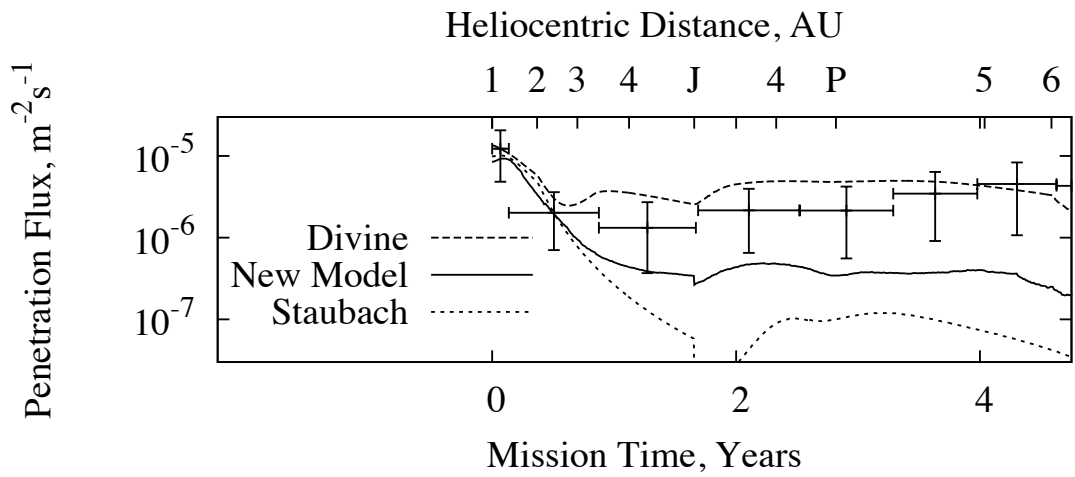,width=0.95\hsize}}
\label{lastpage}
\caption{Fluxes inferred from the Pioneer~11 impact counts
(error bars), and their theoretical counterparts calculated using the Divine,
Staubach, and the new model. Labels on the heliocentric distance
axis are as in Fig.~\ref{p11}.\label{pio11-models}\label{lastfigure}}
\vspace{5cm}
\end{figure}

It was intriguing to compare the performance of all models on those data
that were not incorporated in the new model. Figures~\ref{helios-models}
and \ref{pio11-models} show the three model predictions for the Helios~1
and Pioneer~11 impact records introduced in the previous section.

%
%

The Helios~1 data are described moderately good by both Divine's and new model,
although for the new model it is an extrapolation.
The Staubach model predicts too low a flux, with the ecliptic sensor being dominated
by the interstellar dust.

All three models show a perfect match with the Pioneer~11 data
close to the Earth, however, once the spacecraft leaves the asteroid belt behind,
they start to diverge. The Divine model stays the best description of this data set
up to 6~AU from the Sun where we stop our inspection. In contrast, the Staubach model
falls in significant contradiction with the data that were not the model's base.

In fact, the Divine model explained the Pioneer~11 data by introducing a dedicated
``halo'' population with uniformly distributed orbital planes and low eccentricities.
Although dynamically unstable on the time scale of several tens of Jovian revolutions
about the Sun, this population fits well to the flat flux curve beyond the asteroid belt.

The new model, although being outside the 90\% confidence interval after Jupiter fly-by,
is still capable to reproduce the flat trend with a set of dynamically reasonable populations
of dust from Jupiter-family comets. Lacking the Pioneer~11 data in its base, the new model
shows the strength of the dynamical approach that powers extrapolations, an advantage that
was missing in the previous models.

\section{Conclusions}

A new meteoroid model is developed to predict fluxes on spacecraft
in the Solar system. In this model,
the orbital evolution of meteoroids is for the first
time taken into account in order to complement scarce data.
The Poynting-Robertson effect, and the gravity of planets
are taken into account to produce the orbital distributions of meteoroids
from various known sources in the framework of approximate
analytical solutions. These distributions are fitted
to the infrared observations by the COBE/DIRBE instrument
taken in five wavelength bands from 5 to 100~$\mu$m
over the range of solar elongations from 60$^\circ$ to 130$^\circ$,
the Galileo and Ulysses DDS in-situ flux measurements, and the micro-crater size frequencies
counted on the lunar rock samples retrieved by the Apollo missions.
The wall impacts into the DDS were first taken into account,
the detector's sensitivity to the big impacts was revised downwards
by a factor~16 to reach an agreement with the COBE data.
The AMOR orbital distributions could not be incorporated
in the model since they fall in contradiction with the latitudinal number density
profile of the zodiacal cloud behind the COBE infrared intensities
and some earlier inversions of the zodiacal thermal and visual emissions.
Improvements of the bias correction procedure or model assumptions, such as
the constancy of material density, are required
to fit to the AMOR and COBE data simultaneously.

\bibliography{icarus,dikarev}
\bibliographystyle{icarus}

\newpage

\listoffigures

\pagebreak

\pagebreak

\pagebreak

\pagebreak

\pagebreak

\pagebreak

\pagebreak

\pagebreak

\pagebreak

\pagebreak

\pagebreak

\pagebreak

\pagebreak

\pagebreak

\pagebreak

\pagebreak

\pagebreak

\pagebreak

\pagebreak

\pagebreak

\end{document}